\documentclass[preprint]{revtex4-1}
\usepackage{hyperref}
\usepackage{cleveref}
\usepackage{amssymb}
\usepackage{graphicx}
\usepackage{subfigure}
\usepackage{amsfonts}
\usepackage{amssymb}
\usepackage{latexsym}
\usepackage{natbib}
\usepackage{bm}
\usepackage{newlfont}
\usepackage{epsfig}
\usepackage{ctable}
\usepackage{amsmath}
\usepackage[american]{babel}

\usepackage{placeins}
\usepackage{cleveref}
\crefname{section}{§}{§§}
\Crefname{section}{§}{§§}

\newcommand{\lorenzo}[1]{{\color{black} {#1}}} 
\begin{document}
\title{Suppression of von K\'arm\'an vortex streets past porous rectangular cylinders}
\author{P. G. Ledda$^{1,2}$, L. Siconolfi$^1$, F. Viola$^1$, F. Gallaire$^1$,  S. Camarri$^2$}
\affiliation{$^1$Laboratory of Fluid Mechanics and Instabilities, \'Ecole Polytechnique F\'ed\'erale de Lausanne, Lausanne, CH-1015, Switzerland,\\$^2$Dept. of Industrial and Civil Engineering, Universit\`a di Pisa, Pisa, Italy}

\begin{abstract}
Although the stability properties of the wake past impervious bluff bodies have been widely examined in the literature, similar analyses regarding the flow around and through porous ones are still lacking. In this work, the effect of the porosity and permeability on the wake patterns of porous rectangular cylinders is numerically investigated at low to moderate Reynolds numbers in the framework of direct numerical simulation combined with local and global stability analyses. A modified Darcy-Brinkman formulation \cite{Brinkman1949} is employed here so as to describe the flow behavior inside the porous media, where also the convective terms are retained to correctly account for the inertial effects at high values of permeability. 
Different aspect ratios of the cylinder are considered, varying the thickness-to-height ratios, t/d, from 0.01 (flat plate) to 1.0 (square cylinder). \\
\indent 
The results show that the permeability of the bodies has a strong effect in modifying the characteristics of the wakes and of the  associated flow instabilities, while the porosity weakly affects the resulting flow patterns. In particular, the fluid flows through the porous bodies and, thus, as the permeability is progressively increased, the recirculation regions, initially attached to the rear part of the bodies, at first detach from the body and, eventually, disappear even in the near wakes. {\lorenzo{Global stability analyses lead to the identification of critical values of the permeability above which any linear instability is prevented. Moreover, a different scaling of the non-dimensional permeability allows to identify a general threshold for all the configurations here studied that ensures the suppression of vortex shedding, at least in the considered parameter space.}}
\end{abstract}
\maketitle

\section{Introduction}
\label{sec:intro}

The flow of a liquid phase through a body containing interconnected patterns of voids is frequently encountered in engineering applications as well as in nature. Examples include filtration, where it is necessary to separate solid particles from fluids, cooling systems, where the presence of a porous medium can enhance the heat exchange thus increasing the efficiency, or  water penetration in a sand substrate. Additionally, several minute insects such as the thrips (\textit{Thysanoptera}) use hairy appendages for feeding and locomotion. These filaments-made wings have convenient lift to weight and lift to drag ratios with respect an impervious wing \cite{Sunada2002} and are typically modeled as porous rectangular cylinders \cite{Cummins2017}. In a similar fashion, the seeds of \textit{taraxacum} (commonly known as dandelions) and of \textit{tragopon}, are transported by the wind thanks to a particular umbrella-like extensions called \textit{pappus}, which can be seen as the equivalent of a parachute. Also in this case, the flow pattern past these seeds advected by wind gust can be explained using the model of a porous disk \cite{mcginley1989fruit}, with a Reynolds number based on the pappus diameter around $Re=100$.\\
\indent Inspired by nature and motivated by engineering applications, the fluid dynamics of porous media has received a growing interest over the years. 
At the beginning of the last century, based on the concept that the permeability modifies the flow around a solid object, Prandtl \cite{prandtl1904verhandlungen} designed a passive blowing system to control the flow past a circular cylinder. 
Subsequently, Castro \cite{castro71} studied experimentally the flow around perforated flat plates observing two different flow behaviors: a configuration in which the von~Karman vortex street dominates the wake, and another in which it is inhibited due to the air bleeding from the holes. Furthermore, in some cases, the mean flow is characterized by the presence of a detached recirculation bubble. He also observerd that the transition between these two states, i.e. with attached or detached recirculation region, is quite sudden.

Successively, the turbulent wake past a nominally two-dimensional porous cylinder has been investigated \cite{zong_nepf_2012}, identifying two wake regions: a steady wake region that extends for several cylinder diameters behind the body and a region further downstream associated with the formation of large-scale wake oscillation (von~Karman street). 
Increasing the porosity, and so the permeability, the vortex street formation moves further downstream.
More recently, the problem of the flow around porous square cylinders \cite{jue2004numerical, chen2008numerical} and porous disks \cite{Cummins2017} has been approached numerically. In the latter case, by increasing the disk permeability three different flow regimes have been recognized: (i) first an effectively impervious regime, which is characterized by the presence of a toroidal vortex recirculation region located close to the disk, is observed at low permeability; (ii) Subsequently a transition regime in which the recirculation region shortens and moves downstream occurs for intermediate permeability; (iii) At high permeability regime where the recirculation region is no more present. \\
\indent The scenario described above, however, has been portrayed for the case of steady and stable flow around a porous bluff body at moderate Reynolds numbers, thus overlooking the occurrence of flow instabilities. 
In truth, although the stability properties of the wake past impervious cylinders \cite{MONKEWITZ88,huerre90} and axisymmetric bodies \cite{fabre2008bifurcations, meliga2009unsteadiness} have been extensively investigated in the literature, this is not yet the case for porous objects. 
In this work, the flow patterns around porous rectangular cylinders  and their corresponding stability characteristics have been  investigated systematically for  low-moderate Reynolds numbers by varying the thickness-to-heigh ratio $t$ and the porous medium properties in terms of permeability and porosity.
Firstly, the steady baseflows around permeable rectangular cylinders are computed by solving numerically the incompressible Navier-Stokes equation in the pure fluid domain that are dynamically coupled with a modified Darcy-Brinkman formulation \cite{Brinkman1949}. In this formulation, which has been validated against benchmark results available in the literature, the convective terms are retained to correctly account for the inertia effects at high values of permeability. 
The global stability analysis is then performed as the permeability of the porous medium is progressively changed, finding for each case the marginal stability curve and studying the evolution of the associated eigenvectors. 
The sensitivity of the base flow in different permeability regimes is also evaluated by means of structural sensitivity analysis \cite{Giannetti07}. The regions where the structural sensitivity is stronger define the so called \textit{wavemaker}, where the instability mechanism acts on the baseflow and, with the aim of controlling the instability, which identifies a region where a localized perturbation has an important effect on the eigenvalues.
The onset of a globally unstable mode in the wake of porous cylinders and the possible stabilization effect of the permeability are then connected to the local stability properties of the flow with emphasis on the  relation between the streamwise extension of the absolute region and the one of the recirculation relation. This connection is here explored in detail, showing that global instability can persist even when recirculation are absent in the wake but, nevertheless, the wake velocity defect is sufficiently large. \\
\indent The paper is organized as follows. The flow configuration along with the governing equations for the direct numerical simulations, global and local stability analyses are introduced in section \cref{sec:formulation}.  The numerical method and its validation are detailed in section \cref{sec:numerics}. In \cref{sec:results} the baseflow morphology and its global and local stability properties are first presented as a function of the Reynolds number and permeability. Successively, the effect of porosity and aspect ratios is investigated. Conclusions are outlined in section \cref{sec:results}. Further details on the theoretical formulation and numerical convergence are provided in the appendices \cref{sec:app1} and  \cref{sec:app2}, respectively.

\section{Problem formulation} \label{sec:formulation}

In this section the theoretical framework and the governing equations for the direct numerical simulations, global and local stability analyses are presented.

\subsection{Flow configuration and governing equations}

\begin{figure}[b!]
\begin{center}
\includegraphics[width=.9\textwidth]{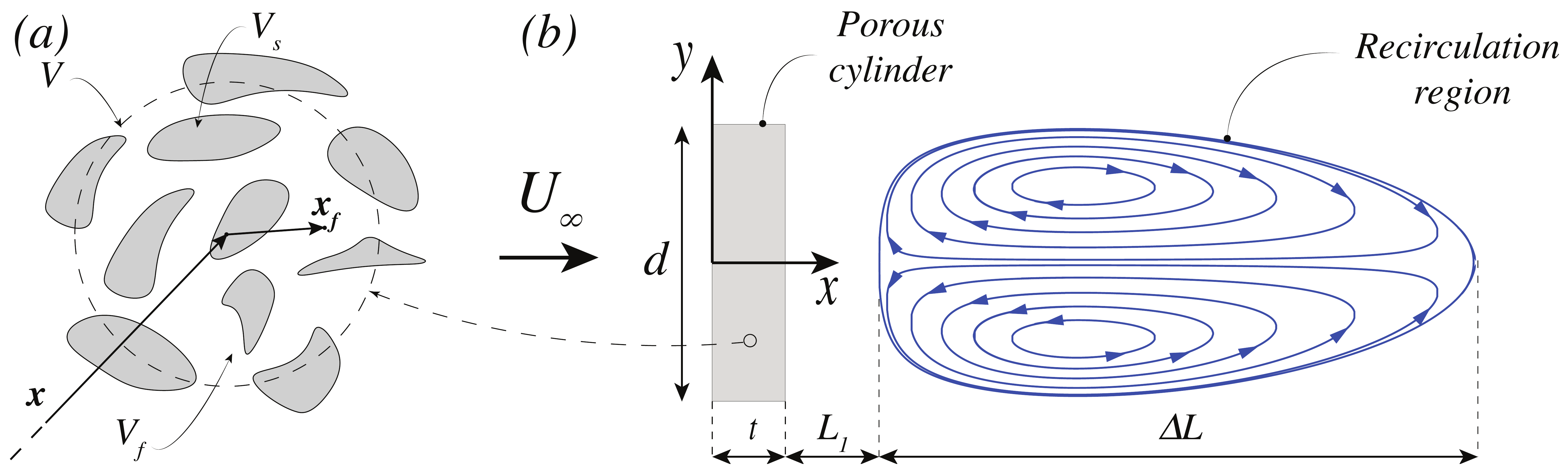}
\caption{(a) Representation of an elementary porous volume $V$. $V_S$ and $V_F$ are the solid and the fluid portions of the volume $V$, respectively. (b) Sketch of the flow configuration: $d$ is the reference dimension of the body, $t$ is its thickness. $L_1$ is the distance between the body and the recirculation bubble when this is detached and $\Delta L$ the length of the recirculation region, both measured on the symmetry line $y=0$.}
\label{fig:flowConf}
\end{center}
\end{figure}
We study the stability of two-dimensional (2D) wakes past porous rectangular cylinders invested by a uniform stream of velocity $U_\infty$ (Fig. \ref{fig:flowConf}) aligned with one of their symmetry axis and orthogonal to their longest sides. The rectangular cylinders, characterized by a thickness-to-height ratio $t/d$, are assumed to be made by a homogeneous and isotropic porous material with porosity $\phi$ and permeability $k$ (as defined in  Appendix~\ref{sec:app1}). The porosity allows the flow to partially pass through the bodies, thus modifying the characteristics of the resulting wakes. It will be shown that the recirculation bubbles can assume different lengths, here labelled as $\Delta L$, depending on the permeability and, for specific configurations, recirculation regions can be detached from body, i.e. $L_1>0$ (see Fig. \ref{fig:flowConf}b). It is important to highlight that, although the present work is focused on the study of the wake past 2D rectangular cylinders, the procedure described in the following can be considered as a general approach to study the instability of the flow past a generic porous bluff body.\\

The fluid motion in the pure fluid region of the domain is described by the velocity field $\mathbf{u}=(u_x,u_y)$ and the pressure field $p$, which satisfy the unsteady incompressible Navier-Stokes equations:
\begin{subequations}
\label{eq:NSdim}
\begin{equation}
\nabla \cdot \textbf{u}=0,
\end{equation}
\begin{equation}
 \rho\bigg(\frac{\partial{\textbf{u}}}{\partial{t}}+\textbf{u}\cdot \nabla  \textbf{u}\bigg) =
 -  \nabla p +\mu \nabla ^2 \textbf{u}
\end{equation}
\end{subequations}
where $\mu$ and $\rho$ are the dynamic viscosity and the density of the fluid, respectively. The set of equations (\ref{eq:NSdim}) are solved in a closed rectangular domain with suitable inlet, lateral and outlet boundaries,  which are specified at the end of this section. 

Concerning the flow inside the body, the porous medium is modelled as a rigid medium completely saturated with fluid. In the literature, different mathematical approaches have been proposed to describe the motion of the fluid inside the pure fluid volume $V_f$ (Fig. 1) of the porous medium \cite{Brinkman1949,bj67,Whitaker1986,Whitaker1996}.
{\lorenzo{In the present work, the approach proposed in \cite{Ochoa-Tapia1995}, which is based on an averaging technique, is adopted. Referring to Fig. 1a, the \textit{superficial volume averaged velocity} can be defined as follows:
\begin{equation}
\langle \textbf{u}_b\rangle |_{\bm{x}}=\frac{1}{V} \int_{V_f} \textbf{u}_b(\bm{x}+\bm{x_f}) \,d\Omega,
\label{superficial}
\end{equation}
where $\bm{x}$ represents the position vector of the centroid of the averaging volume $V$ and $\bm{x_f}$ the position vector of the fluid phase relative to the centroid. Concerning the pressure, it is convenient to define an \textit{intrinsic volume averaged pressure} at the centroid $\bm{x}$ as follows:
\begin{equation}
\langle p_b\rangle_\beta |_{\bm{x}}=\frac{1}{V_f} \int_{V_f} {p}_b(\bm{x}+\bm{x_f}) \,d\Omega,
\label{intrinsic}
\end{equation}
where the average is now made only considering the volume of the fluid phase inside the porous medium. The intrinsic definition of the pressure $\langle p_b\rangle_\beta$, linked with the corresponding superficial one $\langle p_b\rangle$ by the porosity $\phi$, i.e. $\langle p_b\rangle=\phi\langle p_b\rangle_\beta$, results to be convenient since it is a better representation of the pressure that can be measured at the boundary of porous bodies in experiments.
Thus, using the relations (\ref{superficial}) and (\ref{intrinsic}), the average fluid motion inside the porous medium is seen to be governed by the following system of equations:
\begin{subequations}
\begin{equation}
\nabla \cdot \langle \textbf{u}_b\rangle =0,
\end{equation}
\begin{equation}
\frac{\rho}{\phi}\frac{\partial{\langle \textbf{u}_b\rangle }}{\partial{t}}+\frac{\rho}{\phi^2}\langle  \textbf{u}_b\rangle \cdot \nabla \langle  \textbf{u}_b\rangle  =
 -  \nabla\langle  p_b \rangle _{\beta} +\frac{ \mu}{\phi} \nabla ^2\langle  \textbf{u}_b\rangle  -\frac{\mu}{k}  \langle \textbf{u}_b \rangle.
\end{equation}
\end{subequations}} 
Finally, considering $d$ and $U_{\infty}$ as reference length and velocity scales, respectively, the overall system of equations can be written in non-dimensional form as follows:
\begin{itemize} 
\item Pure flow:
\begin{subequations}
\label{DNSfinal1}
\begin{equation}
\nabla \cdot \tilde{\textbf{u}}=0
\end{equation}
\begin{equation}
 \frac{\partial{\tilde{\textbf{u}}}}{\partial{\tilde{t}}}+\tilde{\textbf{u}}\cdot \nabla  \tilde{\textbf{u}} =
 -  \nabla \tilde{p} +\frac{1}{Re} \nabla ^2 \tilde{\textbf{u}}
\end{equation}
\end{subequations}
\item Inside the porous medium:
\begin{subequations}
\label{DNSfinal2}
\begin{equation}
\nabla \cdot \langle\tilde{\textbf{u}}_b\rangle=0
\end{equation}
\begin{equation}
\frac{1}{\phi}\frac{\partial{\langle\tilde{\textbf{u}}_b\rangle}}{\partial{\tilde{t}}}+\frac{1}{\phi^2}\langle\tilde{\textbf{u}}_b\rangle \cdot \nabla \langle \tilde{\textbf{u}}_b\rangle =
 -  \nabla \langle\tilde{p}_b\rangle _{\beta} +\frac{1}{\phi Re} \nabla ^2 \langle\tilde{\textbf{u}}_b\rangle -\frac{1}{ReDa}  \langle\tilde{\textbf{u}}_b\rangle
\end{equation}
\end{subequations}
\end{itemize}
where $Re=\rho U_{\infty} d/\mu$ is the Reynolds number and $Da=k/d^2$ is the Darcy number.
The systems of equations (\ref{DNSfinal1},\ref{DNSfinal2}) are then completed by appropriate boundary conditions. Referring to Fig. \ref{fig:comput}, non-homogeneous Dirichlet boundary conditions specifying the undisturbed incoming flow are applied at the inflow, $\Omega_{in}$, and on the lateral boundaries, $\Omega_{lat}$, i.e. $\tilde{\textbf{u}}=[1,0]$. Stress-free condition is imposed at the outflow boundary, $\Omega_{out}$, i.e. $\textbf{n}\cdot[\mu\nabla\tilde{\textbf{u}}-\tilde{p}\textbf{I}]=0$.
{\lorenzo{Concerning the interface between the cylinder and the outer flow field, the quantities ($\tilde{\textbf{u}},\tilde{p}$) and ($\langle\tilde{\textbf{u}}_b\rangle,\langle\tilde{p}_b\rangle$) can be linked considering that, outside the porous body, the average velocity and pressure correspond, in the present case, to the punctual velocity, i.e. $\langle\tilde{\textbf{u}}\rangle=\tilde{\textbf{u}}$ and $\langle\tilde{p}\rangle=\langle\tilde{p}\rangle_\beta=\tilde{p}$. In particular, assuming a homogeneous porous interface \cite{Ochoa-Tapia1995}, velocity and stress continuity are imposed on $\Omega_{cyl}$, i.e. $\tilde{\textbf{u}}= \tilde{\textbf{u}}_b$ and $\textbf{n}\cdot[\mu\nabla\tilde{\textbf{u}}-\tilde{p}\textbf{I}]=\textbf{n}\cdot[\mu\phi^{-1}\nabla\langle\tilde{\textbf{u}}_b\rangle-\langle\tilde{p}_b\rangle_\beta\textbf{I}]$.
It is important to highlight that the use of different averaging definitions for the velocity (\ref{superficial}) and for the pressure (\ref{intrinsic}) in the continuity of the stresses results to be appropriate thanks to presence of the porosity $\phi$ in the expression of the boundary condition.}} 
In the following, the superscript $~\tilde{\cdot}~$, which indicate non-dimensional quantities, and the average brackets $\langle \cdot \rangle$ will be omitted for sake of brevity. 

\subsection{Global stability and sensitivity analysis}
The occurrence of bifurcations of the system that drive the flow into different flow configurations is studied in the framework of linear stability analysis. Using a unified nomenclature for the flow field inside and outside the porous body, (u,p) we consider the flow solution as the superposition of a steady {\textit{baseflow}} $(\textbf{U},P)(x,y)$ and an infinitesimal unsteady perturbation $(\textbf{u}',p')(x,y,t)$. As concerns the flow description outside the porous body, introducing this decomposition in the system (\ref{DNSfinal1}), two mathematical problems are obtained describing the spatial structure of the baseflow and the evolution of the unsteady perturbations. The baseflow outside the porous body is governed by the steady version of the system (\ref{DNSfinal1}). Perturbations of the baseflow are sought in the form of normal modes, i.e. $(\textbf{u}',p')(x,y,t)=(\hat{\textbf{u}},\hat{p})(x,y)~\mathrm{exp}(\sigma t)$, where $\sigma$ is the eigenvalue associated with the corresponding eigenfunction  $(\hat{\textbf{u}},\hat{p})(x,y)$. The dynamics of an infinitesimal perturbation can be then described by the unsteady Navier-Stokes equations, linearized around the baseflow solution ($\textbf{U},P$), that can be written as:
\begin{subequations}
\label{LNSfinal1}
\begin{equation}
\nabla \cdot \hat{\textbf{u}}=0
\end{equation}
\begin{equation}
 \sigma \hat{\textbf{u}}+\textbf{U}\cdot \nabla  \hat{\textbf{u}}+\hat{\textbf{u}}\cdot \nabla \textbf{U}  =
 -  \nabla \hat{p}+\frac{1}{Re} \nabla ^2 \hat{\textbf{u}}.
\end{equation}
\end{subequations}
The same procedure can be applied disturbance dynamics inside the porous medium. The baseflow inside the body is given by the steady version of system (\ref{DNSfinal2}), while the perturbation dynamics is given by:
\begin{subequations}
\label{LNSfinal2}
\begin{equation}
\nabla \cdot \hat{\textbf{u}}=0
\end{equation}
\begin{equation}
\frac{1}{\phi}\sigma \hat{\textbf{u}}+\frac{1}{\phi^2}(  \textbf{U}\cdot \nabla\hat{\textbf{u}}+\hat{\textbf{u}}\cdot \nabla \textbf{U})=
-  \nabla \hat{p} +\frac{1}{\phi Re} \nabla ^2 \hat{\textbf{u}}  -\frac{1}{ReDa}  \hat{\textbf{u}}
\end{equation}
\end{subequations}

The linearized systems (\ref{LNSfinal1},\ref{LNSfinal2}) are then completed with the following boundary conditions: homogeneous Dirichlet condition is imposed at the inlet $\Omega_{in}$ and on the lateral boundaries of the domain $\Omega_{lat}$, while the stress-free condition is considered at the outflow $\Omega_{out}$. At the fluid-porous interface $\Omega_{cyl}$, velocity and stress continuity condition between inner and outer disturbances are applied in similarity of the interfacial condition imposed in the systems (\ref{DNSfinal1},\ref{DNSfinal2}).

The systems (\ref{LNSfinal1},\ref{LNSfinal2}), together with their boundary conditions, define an eigenvalue problem with, possibly, complex eigenvalues $\sigma_n=\lambda_n + i ~\omega_n$. The real part of the eigenvalue, $\lambda_n$, is the growth rate of the global mode, whereas the imaginary part, $\omega_n$, is its angular velocity. Thus, sorting the eigenvalues by their growth rates in descending order, i.e. $\lambda_0>\lambda_1>\lambda_2,...$, the system is considered asymptotically stable if the growth rate of the leading eigenvalue $\lambda_0$ is positive, while it is asymptotically unstable if $\lambda_0$ is negative.

Following \cite{Giannetti07}, the evaluation of the sensitivity of the global eigenvalue to a structural perturbation $\delta\mathcal{L}$ of the linear operators of the systems (\ref{LNSfinal1},\ref{LNSfinal2}) allows to highlight the region of the flow field where the instability mechanism acts on the baseflow. In particular, considering a localized forcing $\textbf{f}(x_0,y_0)=\mathbf{A_0} \hat{u}(x,y) \delta(x-x_0,y-y_0)$ acting on equations (6,7), where $\mathbf{A_0}$ is a generic feedback matrix and $\delta$ is the 2D Dirac function, the induced eigenvalue variation $\delta \sigma$ can be maximized as follows:
\begin{equation}
|\delta \sigma (x_0,y_0)| \leq ||\mathbf{A_0}|| \cdot || \hat{\textbf{u}}(x_0,y_0)|| \cdot ||\hat{\textbf{u}}^{\dag}(x_0,y_0)||
\end{equation}
where $\hat{\textbf{u}}^{\dag}$ is the adjoint velocity field, solution of the following system of adjoint equations:
\begin{itemize}
\item In the clear fluid:
\begin{subequations}
\label{ALNS1}
\begin{equation}
\nabla \cdot \hat{\textbf{u}}^{\dag}=0
\end{equation}
\begin{equation}
 \sigma \hat{\textbf{u}}^{\dag}-\textbf{U}\cdot \nabla  \hat{\textbf{u}}^{\dag}+\hat{\textbf{u}}^{\dag}\cdot (\nabla \textbf{U})^T =
 -  \nabla \hat{p}^{\dag} +\frac{1}{Re} \nabla ^2 \hat{\textbf{u}}^{\dag}
\end{equation}
\end{subequations}
\item In the porous medium:
\begin{subequations}
\label{ALNS2}
\begin{equation}
\nabla \cdot \hat{\textbf{u}}^{\dag}=0
\end{equation}
\begin{equation}
\frac{1}{\phi}\sigma \hat{\textbf{u}}^{\dag}+\frac{1}{\phi^2}( - \textbf{U}\cdot \nabla\hat{\textbf{u}}^{\dag}+\hat{\textbf{u}}^{\dag}\cdot (\nabla \textbf{U})^T  )=
-  \nabla \hat{p}^{\dag} +\frac{1}{\phi Re} \nabla ^2 \hat{\textbf{u}}^{\dag}  -\frac{1}{ReDa}  \hat{\textbf{u}}^{\dag}.
\end{equation}
\end{subequations}
\end{itemize}
The boundary conditions that complete the systems (\ref{ALNS1},\ref{ALNS2}) are the same used for the direct problem defined in the systems (\ref{LNSfinal1},\ref{LNSfinal2}), except for the outflow boundary condition that can be written as $\textbf{n}\cdot\left(Re^{-1}\hat{\textbf{u}}^{\dag}- \hat{p}^{\dag}\textbf{I}\right)=-\left(\textbf{U} \cdot \textbf{n}\right)\hat{\textbf{u}}^{\dag}$.

\subsection{Spatio-temporal stability analysis}
As mentioned before, the properties of the porous bodies affect the characteristics of the wake and therefore its stability. In order to analyze in detail the nature of the instability and its changes with the porosity and the permeability of the body, a spatio-temporal stability analysis is carried out. 
Under the assumption of weakly non-parallel flow, the velocity profile at each streamwise section is extracted and its stability is studied inspecting the growth rate of a local perturbation of the form  $\textbf{u}^{\prime}=\textbf{u}^{*}\exp\left[i(k x -\omega t)\right]$, where $k$ is the local wavenumber and $\omega$ the angular velocity. In order the study the absolute and convective nature of the stability, the Briggs-Bers method is used \cite{huerre90}, which consists in defining the saddle point $k_0$ in the complex $k$-space, i.e. $\partial{\omega}/\partial{k}(k=k_0)=0$ (the \textit{absolute wavenumber}). If the imaginary part of the corresponding absolute frequency $\omega_0$ is greater then zero, i.e. $Im({\omega_0})>0$, the flow profile is \textit{absolutely unstable}, otherwise it is convectively unstable. As pointed out in literature \cite{huerre90,Gallaire2017}, the extension of the absolute unstable region can be linked to the characteristics of the global stability of the wake.

\section{Numerical method}\label{sec:numerics}
\label{sec:numerical}
\begin{figure}[t!]
\begin{center}
\includegraphics[width=.9\textwidth]{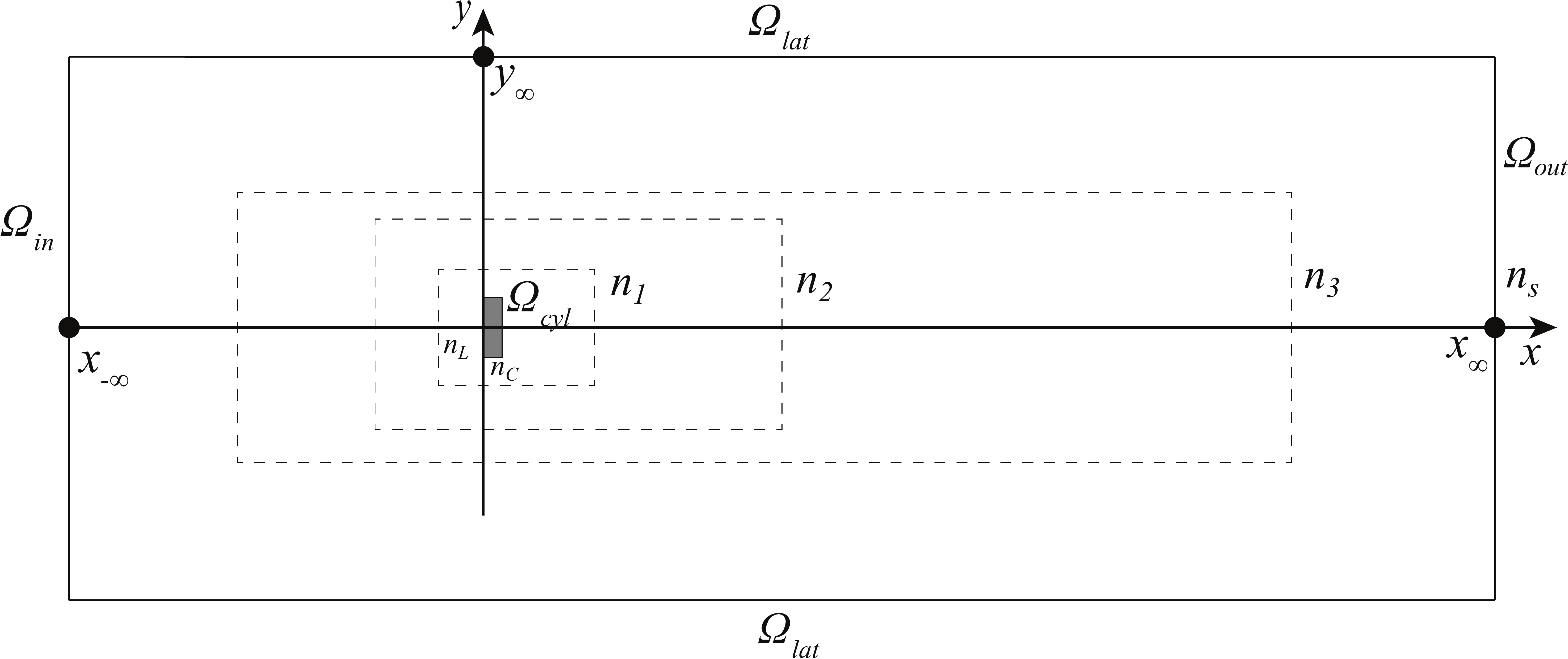}
\caption{Sketch of the computational domain. The porous cylinder corresponds to the gray area whereas the dashed line rectangles depict the mesh refinement regions. The spatial extent of the computational domain is defined by the location of the boundaries
$x_{-\infty}$,  $x_{+\infty}$ and $y_{\infty}$ and the level of the mesh refinement is controlled by the vertex densities $n_L$, $n_C$ $n_1$, $n_2$, $n_3$ and $n_s$.}
\label{fig:comput}
\end{center}
\end{figure}

In this section, the numerical methods employed to solve the governing equations introduced in Section \ref{sec:formulation} are described. 
The evaluation of the baseflow, steady solution of the systems of equations (\ref{DNSfinal1},\ref{DNSfinal2}), and the solution of the global eigenvalue problems (\ref{LNSfinal1},\ref{LNSfinal2}) over the rectangular domain sketched in Fig. \ref{fig:comput} are carried out using FreeFem++ \cite{MR3043640} solver. The spatial discretisation is then obtained by a finite-element formulation based on the Taylor-Hood elements.  The unstructured grid is made of five regions of refinement (see Fig. \ref{fig:comput}), where the vertex densities have been chosen  after a convergence study, whose results are reported in detail in Appendix B.

As regards the identification of the absolute and  spatio-temporal analysis, the 1D velocity profile extracted form  the baseflow at several streamwise positions is considered parallel as in the local stability analysis \cite{huerre90}. The resulting parallel linear equations are then discretized using a pseudo-spectral method employing Gauss-Lobatto-Legendre collocation points. The saddle-point in the complex wavelength space $k$ of the local angular velocity $\omega$ is then localized using a Newton iterative method.
\subsubsection{Validation of the model and its  implementation against the literature}
In this section, the mathematical and numerical approaches described in Sec.(\ref{sec:formulation}) and (\ref{sec:numerical}) are validated against the results reported in \cite{zampogna16}. The test case consists in the DNS of the 2D flow in a square cavity of dimension $L$, where homogeneous Dirichlet boundary conditions are applied at all the boundaries except for the top one, where a uniform tangential velocity, i.e. $u_x=\bar{U}$, is considered. For a height ($y-$direction) of $0.33L$ the cavity is occupied by a porous medium, whose porosity is fixed at $\phi=0.8$ and its permeability in the $x-$ and $y-$direction is $k=1.052 \times 10^{-5}$ and $k=2.196 \times 10^{-5}$, respectively.
Exploiting the formulations described in equations (\ref{DNSfinal1}) and (\ref{DNSfinal2}), direct numerical simulations (DNSs) are carried out at $Re_L={\rho\bar{U}L}/{\mu}=100$ and the results are reported in Fig. \ref{fig:valid}, together with the reference ones from \cite{zampogna16}, which are reported as red points. A good agreement has been found between the two sets of data, especially at the fluid-porous interface, where the velocity gradients are correctly estimated. This good agreement results from having retained the convective terms in the equations (\ref{DNSfinal2}), that allows the inertial effects to penetrate inside the porous medium, according to the discussion of \cite{zampogna16}. In  summary, since the DNS data in \cite{zampogna16} have been obtained using a numerical method and resolution which are different and independent from the ones adopted here, we can state that results in Fig. \ref{fig:valid} validate (i) the model employed for the porous media, (ii) the interface boundary conditions between the porous body and the external flow and, lastly, (iii) the numerical implementation of the proposed sets of equations.

\begin{figure}[h!]
\includegraphics[width=\textwidth]{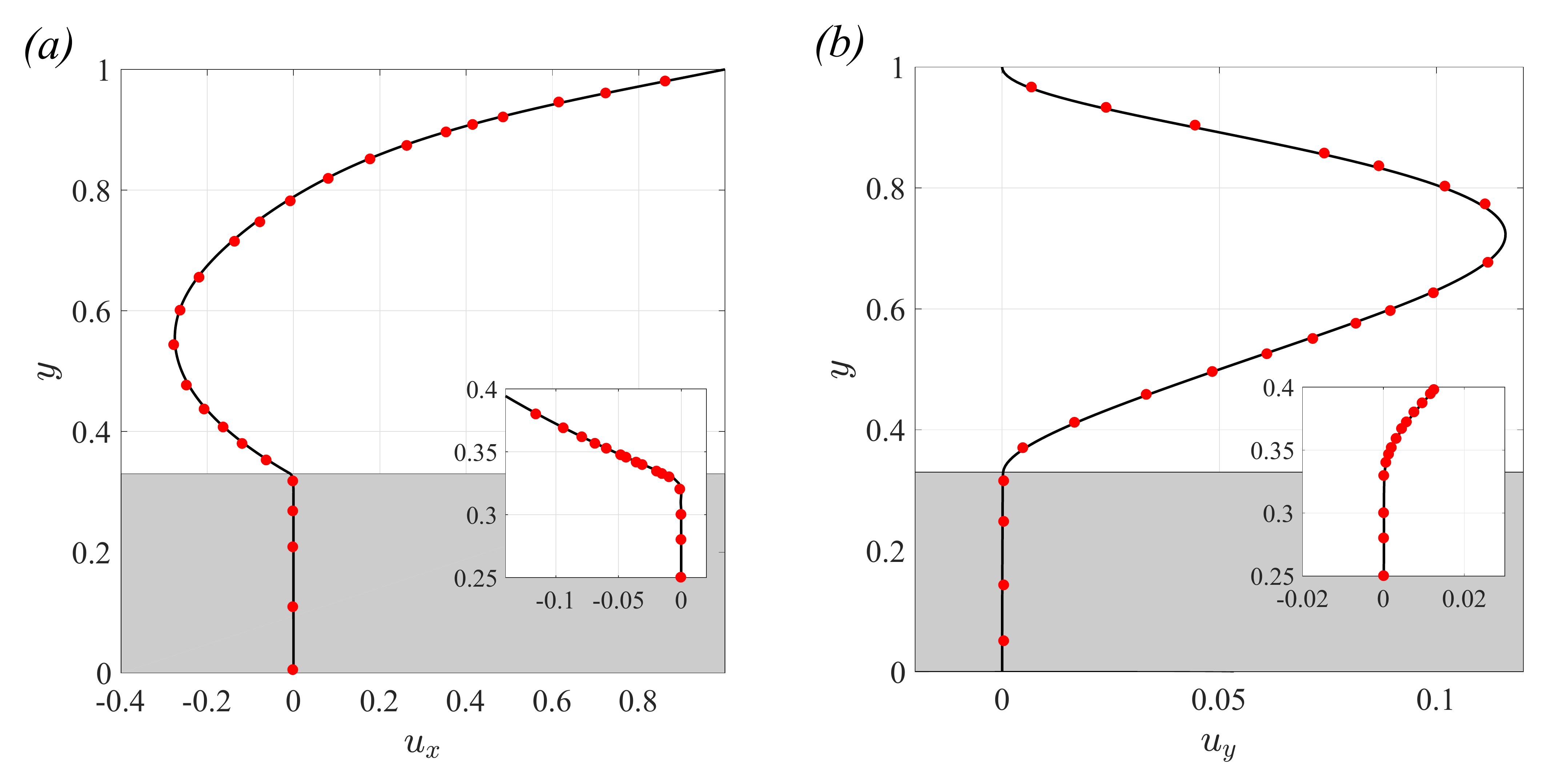}
\caption{Velocity profiles (solid lines) of (a) $u_x$ and (b) $u_y$ extracted at half of the cavity, i.e. $x=0.5$, compared to the results of \cite{zampogna16} (red dots).}
\label{fig:valid}
\end{figure}

\section{Results}\label{sec:results}
In this section, the results of the present work in terms of base flow characterization and stability analysis are described in detail. In particular, different rectangular cylinders are considered, varying the thickness-to-height ratio $t/d$ from $0.01$, i.e. a flat plate in good approximation, to $1.0$, i.e. a square cylinder. We anticipate that the baseflow morphology and stability properties of the flow weakly depend on the ratio 
$t/d$ and the porosity $\phi$. 
For this reason, we first focus on the the effect of the permeability, $k$, and of the Reynolds number, $Re$, on the flow field in the case $t/d=0.25$ and $\phi=0.65$.
Subsequently, the effect of the aspect ratio, $t/d$ and of the porosity, $\phi$ on the results will be discussed.

\subsection{Rectangular cylinder with t/d=0.25}
\subsubsection{Base flow}
The base flow consists, for all the Reynolds numbers here considered, in two perfectly symmetric and counter-rotating recirculation bubbles located in the wake of the cylinder. The geometric characteristics of the recirculation regions depend, however, on the considered Reynolds number, $Re$, and on the permeability, $k$, of the body. In particular, let us first study the effect of the permeability, keeping fixed the Reynolds number at $Re=30$. At low values of the permeability, e.g. $Da\approx10^{-10}$, the resulting flow field is very similar to the one that occurs around a solid cylinder, where the recirculation bubbles lie in the near wake of the cylinder and remain attached at its base (see Fig. \ref{fig:streamDaVar}a). Increasing the permeability, i.e. increasing the Darcy number, the flow field inside the cylinder becomes not negligible and, for a critical value of the Darcy number, $Da_{cr1}$, the recirculation bubbles detach from the base of the cylinder, as visible form the streamline patterns reported in Fig. \ref{fig:streamDaVar}b,c. Finally, further increasing the Darcy number, a second critical value is present, $Da_{cr2}$, such that recirculation regions disappear, as shown in Fig. \ref{fig:streamDaVar}d and only a wake velocity defect is present past the cylinder.  
In particular, for $Re=30$ the critical values are $Da_{cr1}=1\times10^{-7}$ and $Da_{cr2}=1.5\times10^{-3}$.

\begin{figure}[h!]
\begin{center}
\includegraphics[trim=0mm 0mm 15mm 50mm,clip,width=.495\textwidth]{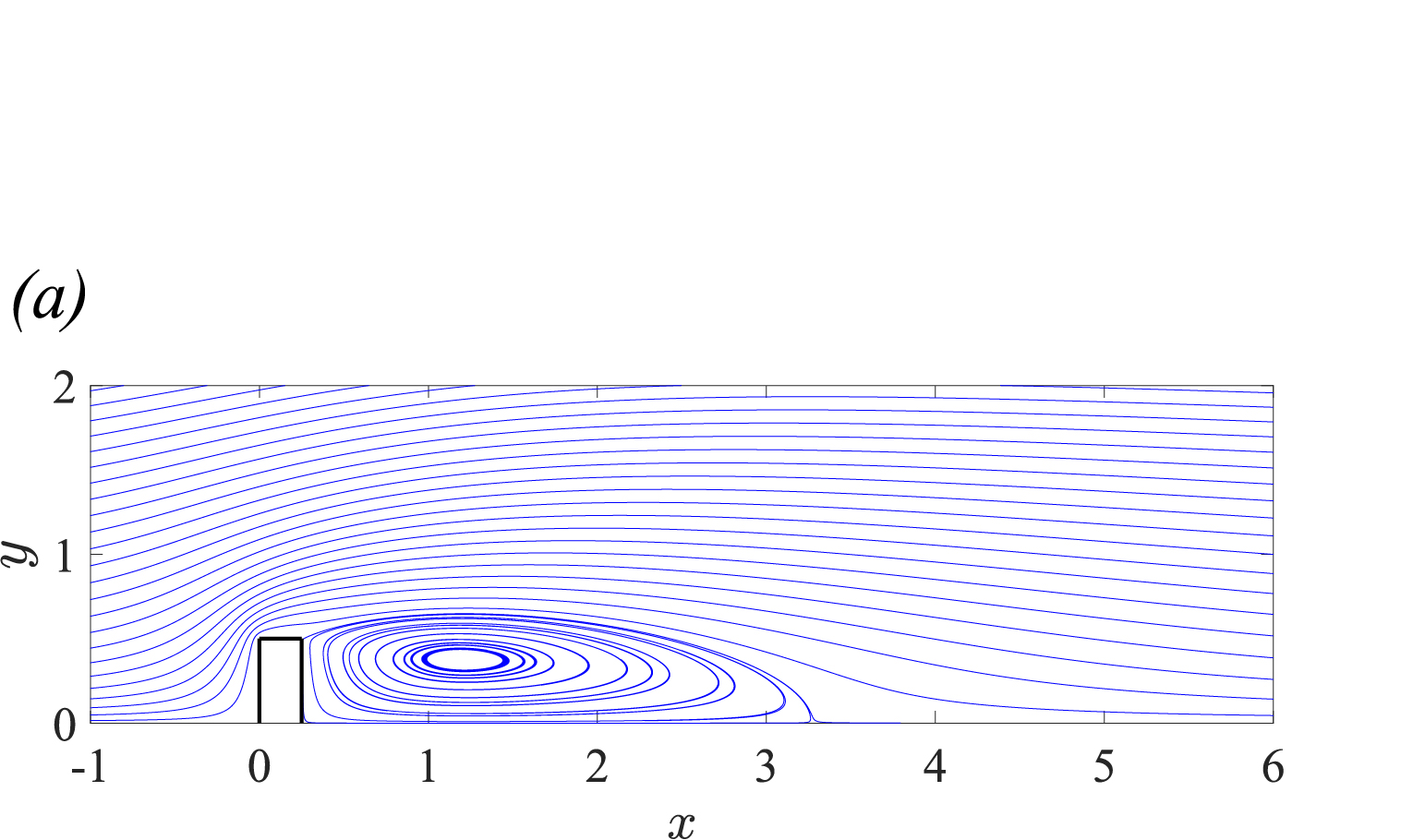}
\includegraphics[trim=0mm 0mm 15mm 50mm,clip,width=.495\textwidth]{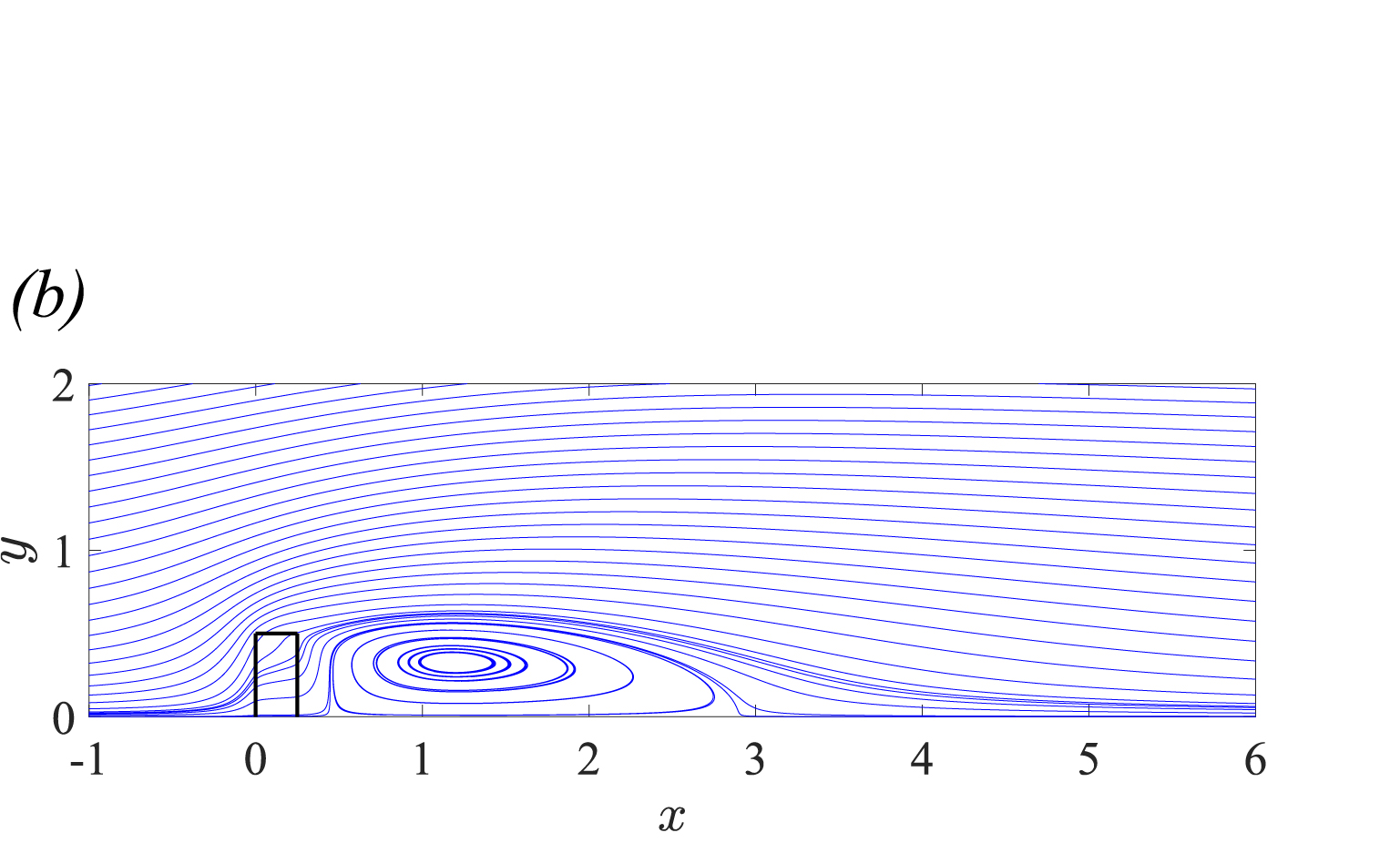}\\
\includegraphics[trim=0mm 0mm 15mm 40mm,clip,width=.495\textwidth]{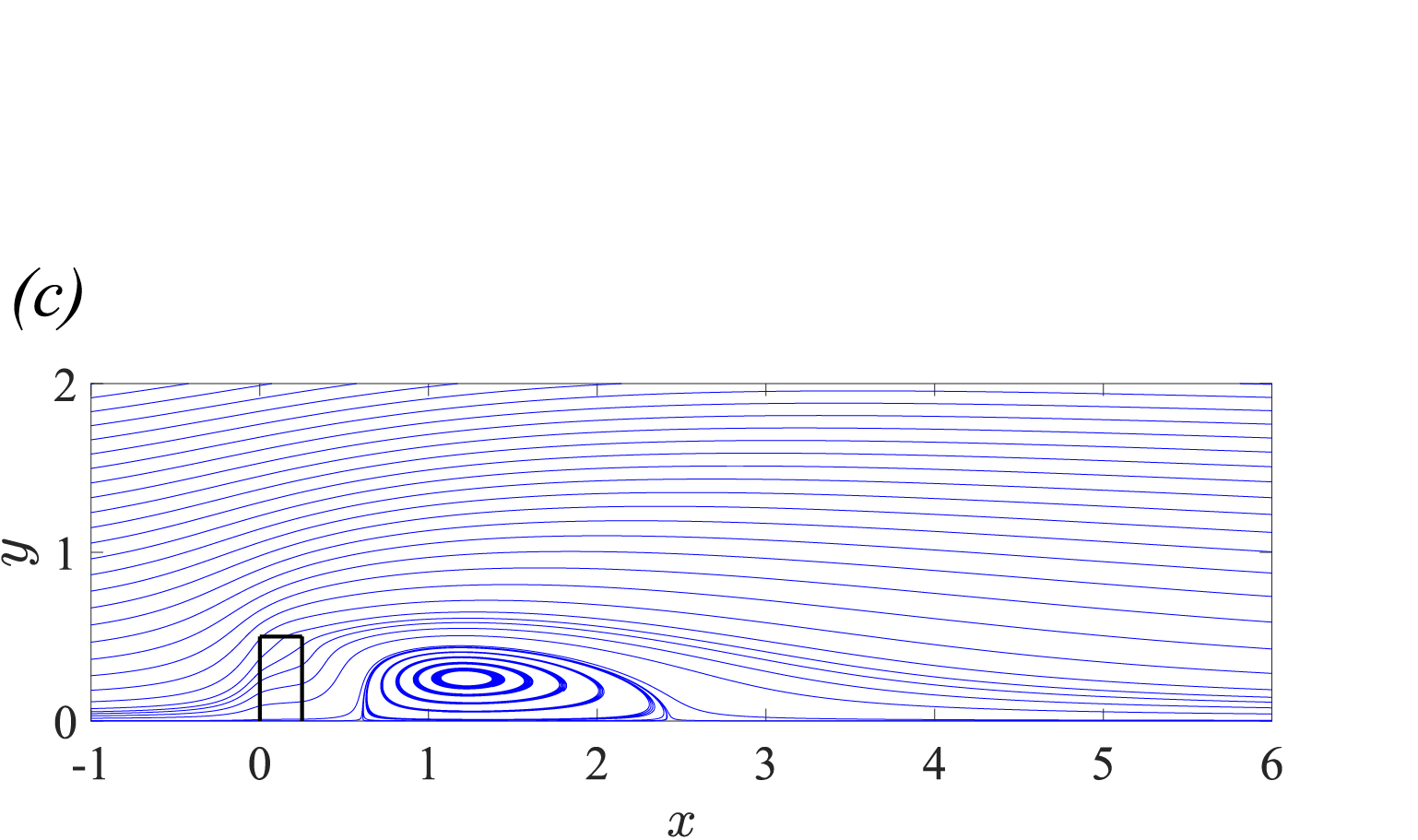}
\includegraphics[trim=0mm 0mm 15mm 40mm,clip,width=.495\textwidth]{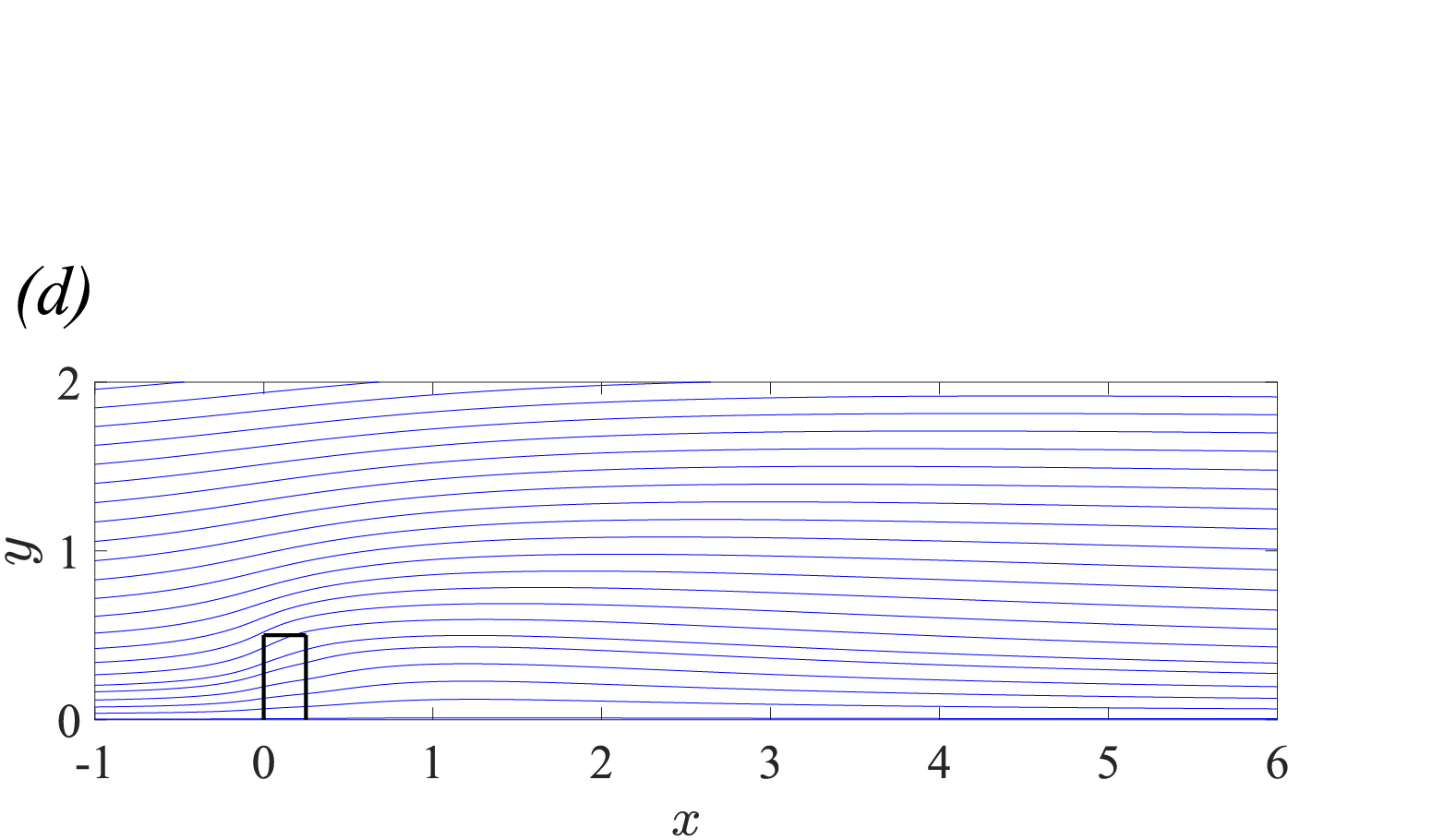}\\
\caption{Baseflow. Flow streamlines at $Re=30$, and (a) $Da=10^{-10}$, (b) $Da=5 \times 10^{-4}$, (c) $Da=1.1 \times 10^{-3}$, (d) $Da=5 \times 10^{-3}$ (only half of the domain, i.e. $y\ge 0$, is shown).}
\label{fig:streamDaVar}
\end{center}
\end{figure}
\FloatBarrier

As visible in Fig. \ref{fig:streamDaVar}, the streamlines at the upper,  i.e. $x=0$ and $y=0.5$, and lower corners, i.e. $x=t$ and $y=0.5$, of the cylinder are modified, due to the characteristics of the material that allows the flow to pass through the body (see Fig. \ref{fig:vectorsSolidPorous}). In particular, the resulting shapes and dimensions of the recirculation bubbles can be related, then, to the vorticity field. Increasing the permeability, the intensity of the vorticity at the two separation points decreses. This is clearly visible in Fig. \ref{fig:vort2}, where the colour-contours of vorticity in the neighbourhood of the upper corner are depicted.
The less intense vorticity observed in the porous case leads to a reduction of the induced counter velocity in the wake, leading to a smaller recirculation length in the streamwise direction and lower backwards velocity intensity. Moreover, since the two vorticity layers are closer to the centre line as the $Da$ number is increased, a consequent reduction of the width of the recirculation bubbles in $y-$direction is also found (see Fig. \ref{fig:streamDaVar}). \\

\begin{figure}[h!]
\begin{center}
\includegraphics[width=1\textwidth]{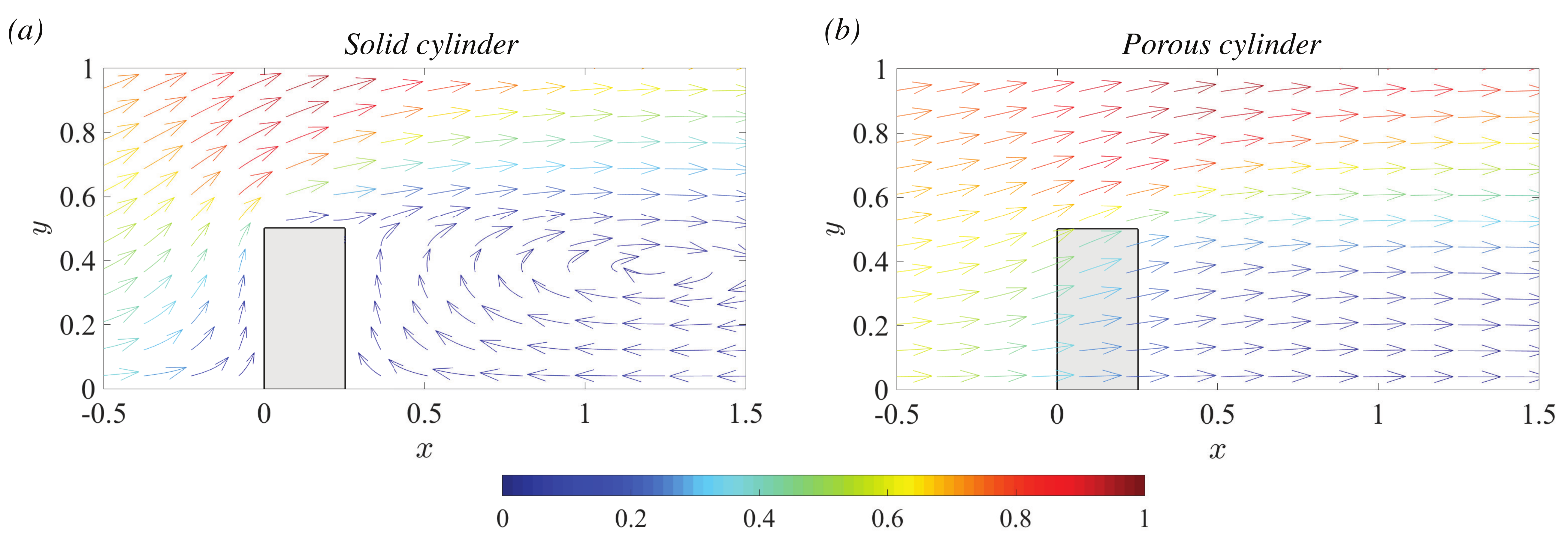}
\caption{Representative behaviour of the velocity field for the flow past a (a) solid and (b) porous rectangular cylinder ($Da=5 \times 10^{-3}$) at $Re=30$. The colours represents the velocity magnutude.}
\label{fig:vectorsSolidPorous}
\end{center}
\end{figure}

\begin{figure}[h!]
\begin{center}
\includegraphics[trim=0mm 0mm 15mm 0mm,clip,width=.325\textwidth]{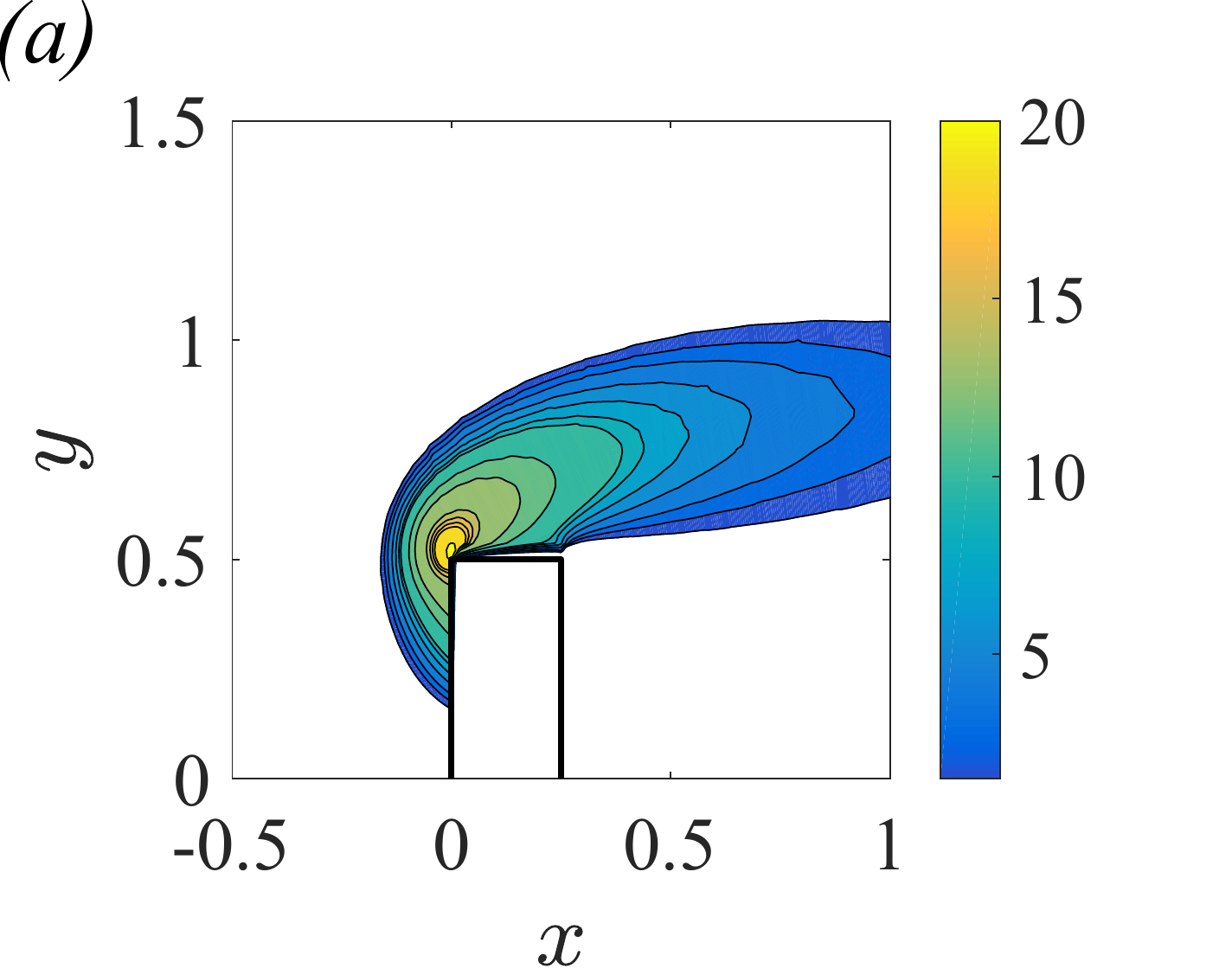}
\includegraphics[trim=0mm 0mm 15mm 0mm,clip,width=.325\textwidth]{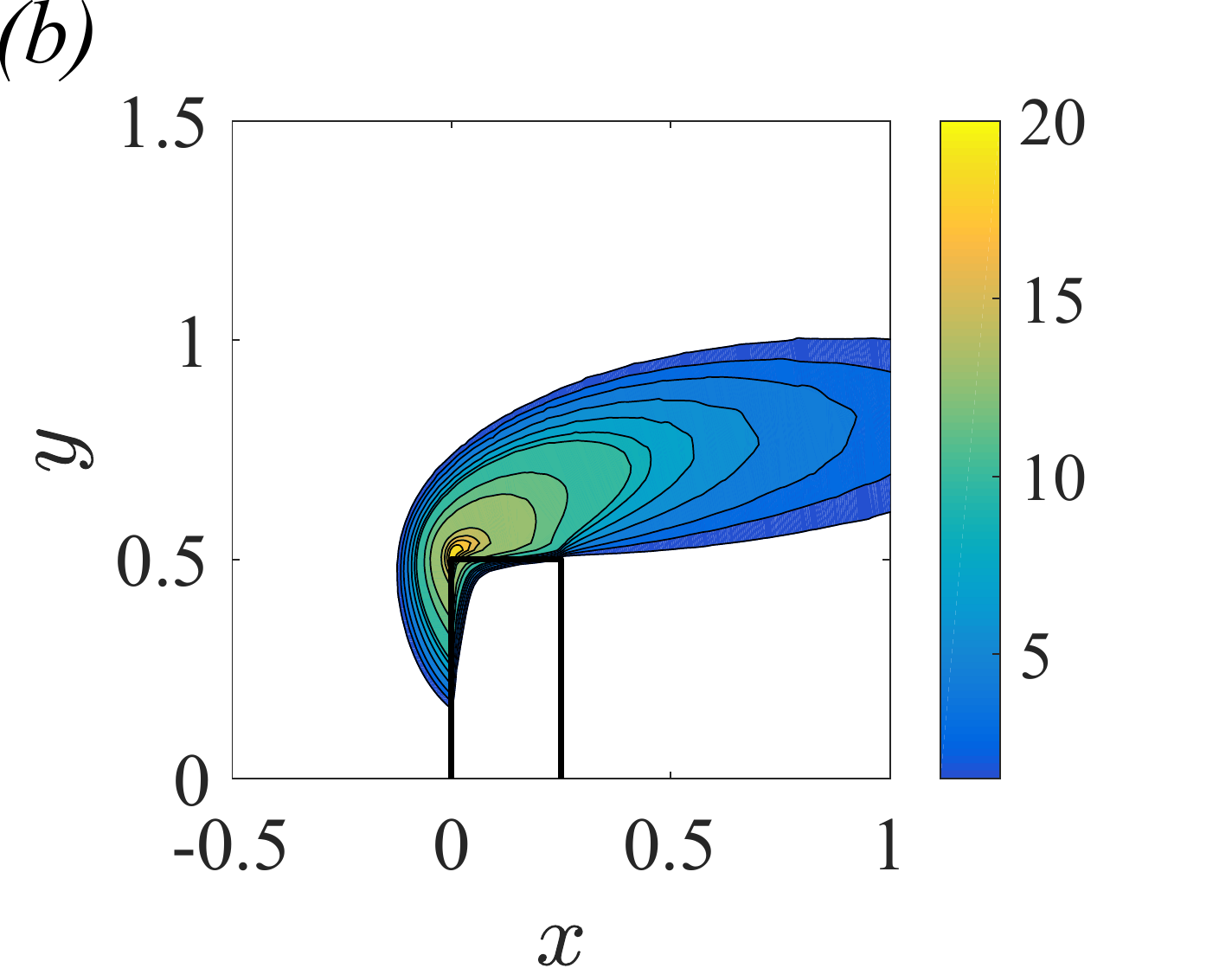}
\includegraphics[trim=0mm 0mm 15mm 0mm,clip,width=.325\textwidth]{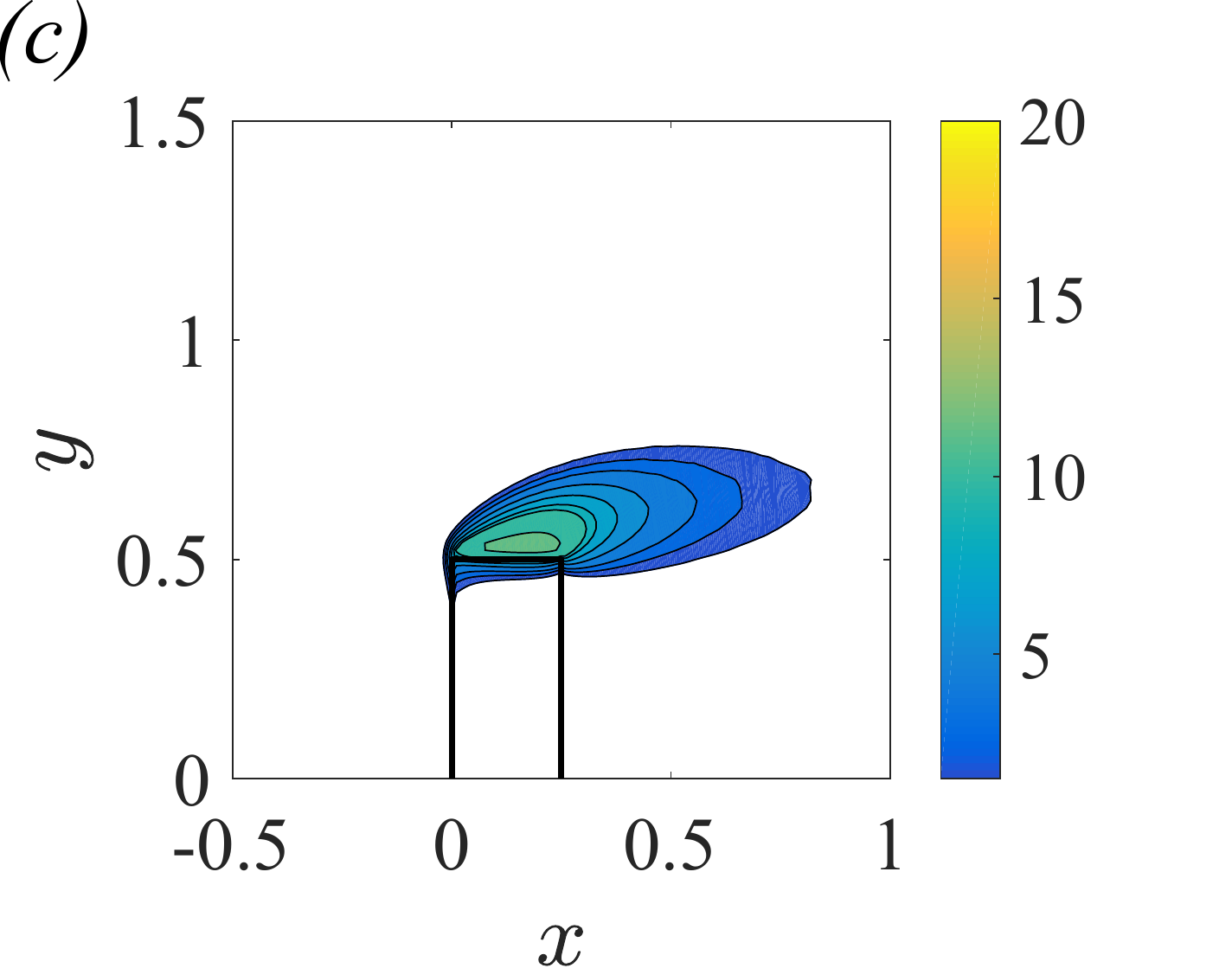}
\caption{Distribution of vorticity in the upper part of the body, at $Re=30$ and (a) $Da=10^{-10}$, (b) $Da=5 \times 10^{-4}$, (c) $Da=5 \times 10^{-3}$. }
\label{fig:vort2}
\end{center}
\end{figure}
\FloatBarrier

Similar effects are also found by fixing the permeability and increasing the flow Reynolds number. Fig. \ref{fig:flowConf1} shows the streamline patterns for different values of the Reynolds number with constant permeability and porosity set to $Da=1.1 \times 10^{-3}$ and $\phi=0.65$, respectively. It is possible to observe that the streamwise extension of the recirculation region, as the flow Reynolds number is increased, increases at first, successively decreases and finally disappears. At the same time, once detached, the recirculation bubble gets progressively more distant from the cylinder, i.e. $L_1$ increases monotonically with $Re$. The dependence of the recirculation bubble length on the flow Reynolds number depends on the competition between the inertia terms and the Darcy terms in the systems (\ref{DNSfinal1}) and (\ref{DNSfinal2}). As the flow Reynolds number is increased, the inertial terms become more important and, at the same time, the viscous drag effects inside the porous media are reduced. Thus, the fluid can easier pass through the body, strongly modifying the velocity field and, as a consequence, the vorticity field. 

In particular, when $Re$ is increased, the generated vortical structures become more intense (see \ref{fig:vort3}b,c), with a consequent elongation of the recirculation bubbles. Successively, further increasing the Reynolds number, the vorticity magnitude decreases (as in \ref{fig:vort3}c), yielding the recirculation bubbles to shorten until they disappear. 

\begin{figure}[h!]
\begin{center}
\includegraphics[trim=0mm 0mm 15mm 50mm,clip,width=.495\textwidth]{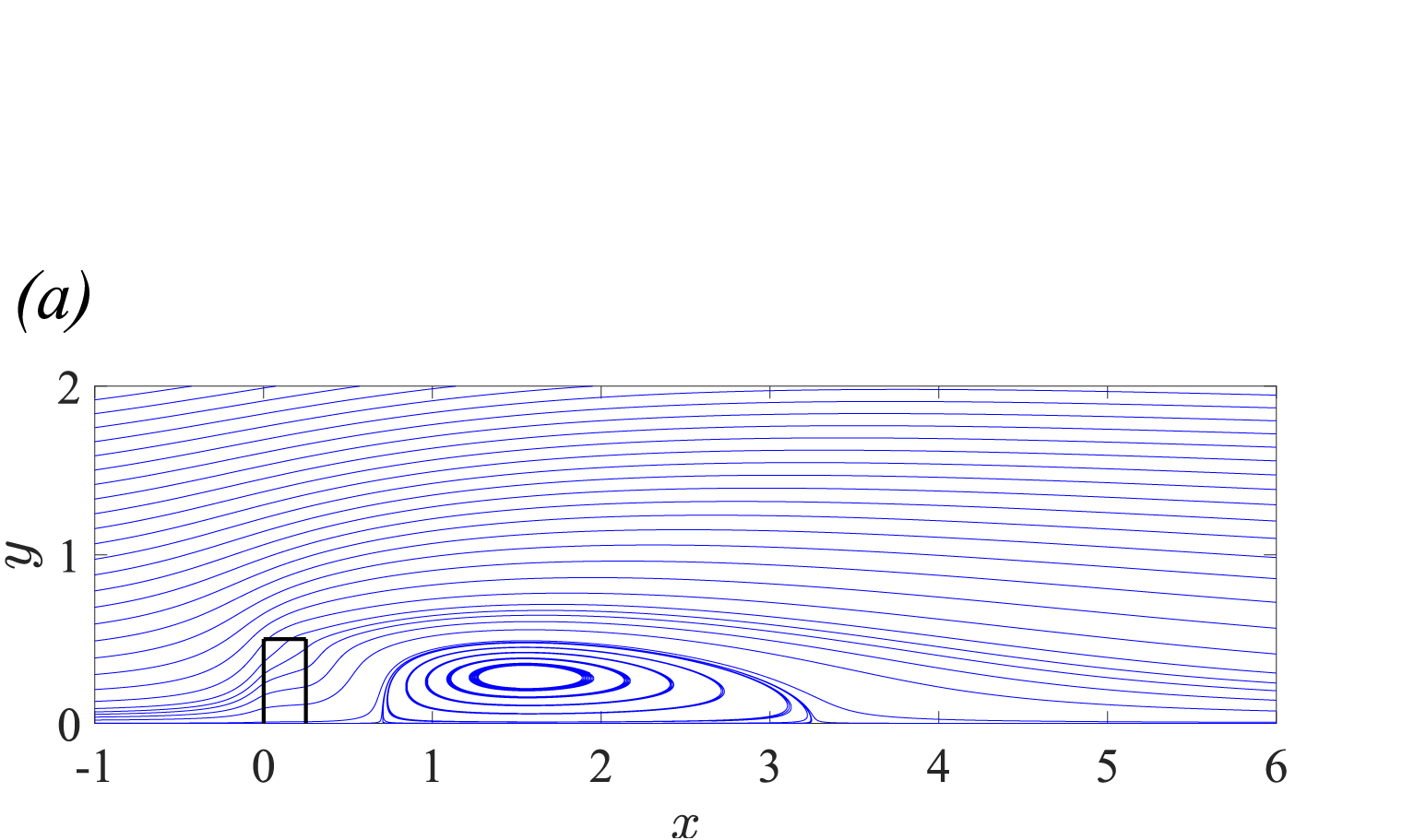}
\includegraphics[trim=0mm 0mm 15mm 50mm,clip,width=.495\textwidth]{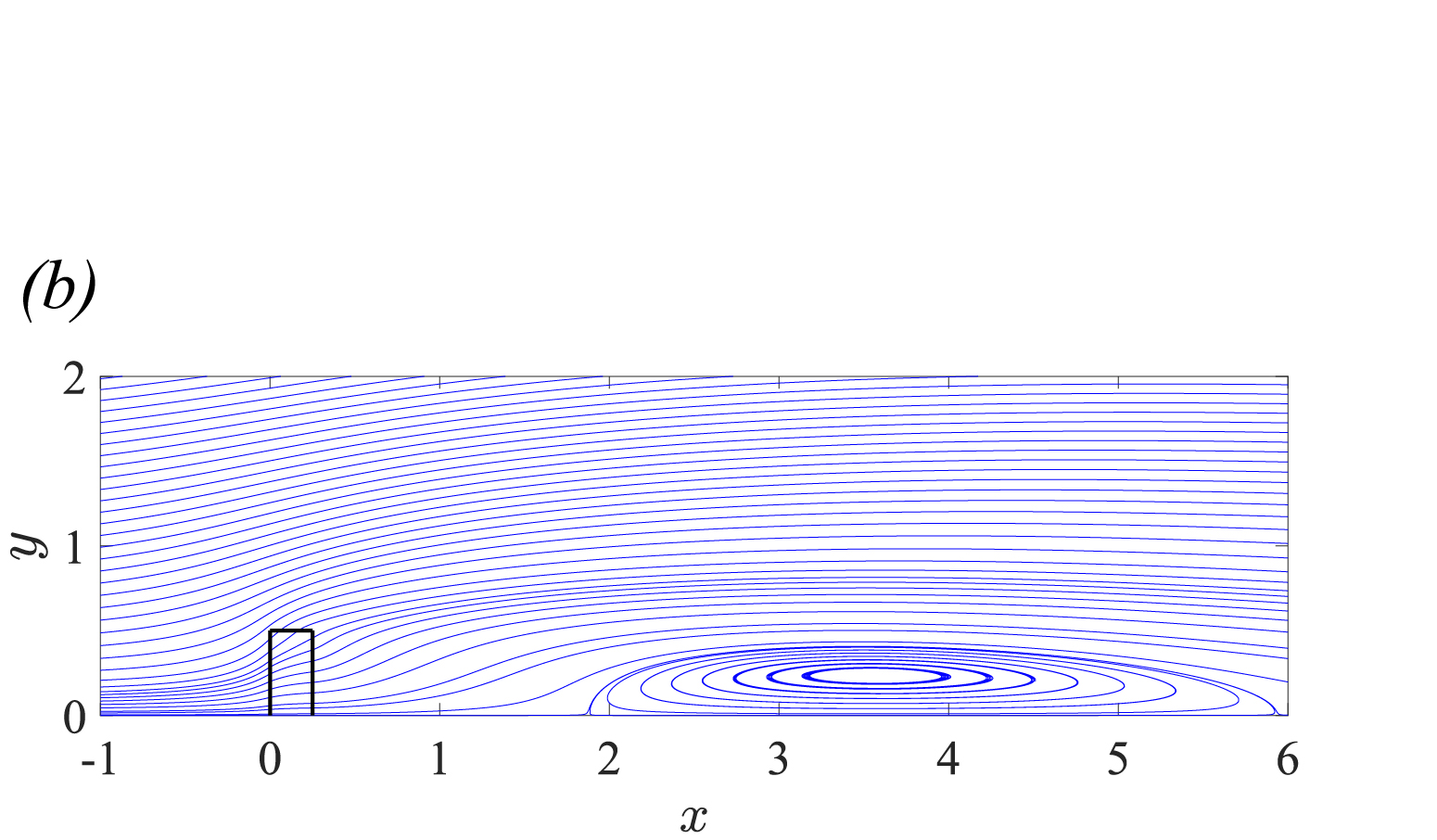}\\
\includegraphics[trim=0mm 0mm 15mm 50mm,clip,width=.495\textwidth]{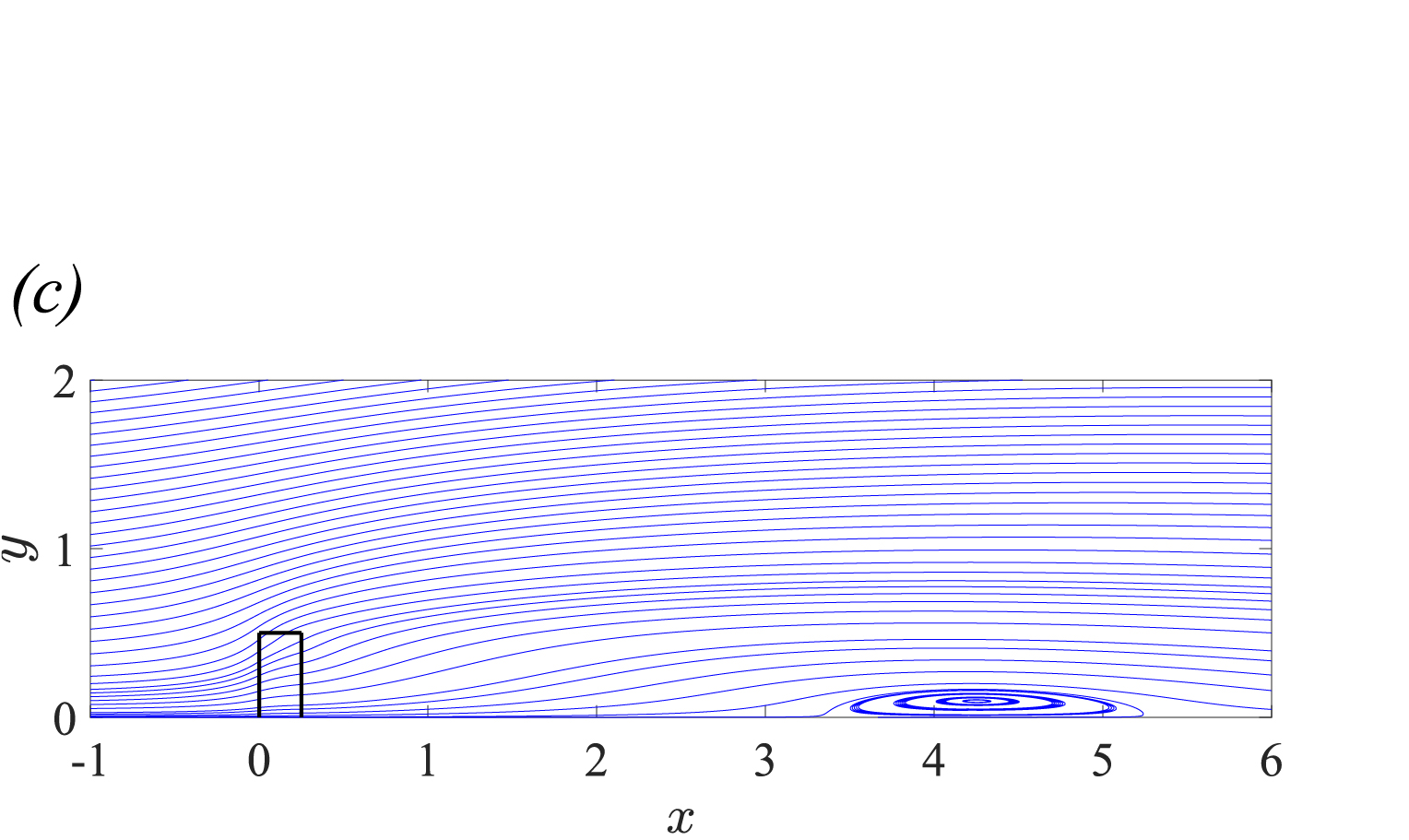}
\includegraphics[trim=0mm 0mm 15mm 50mm,clip,width=.495\textwidth]{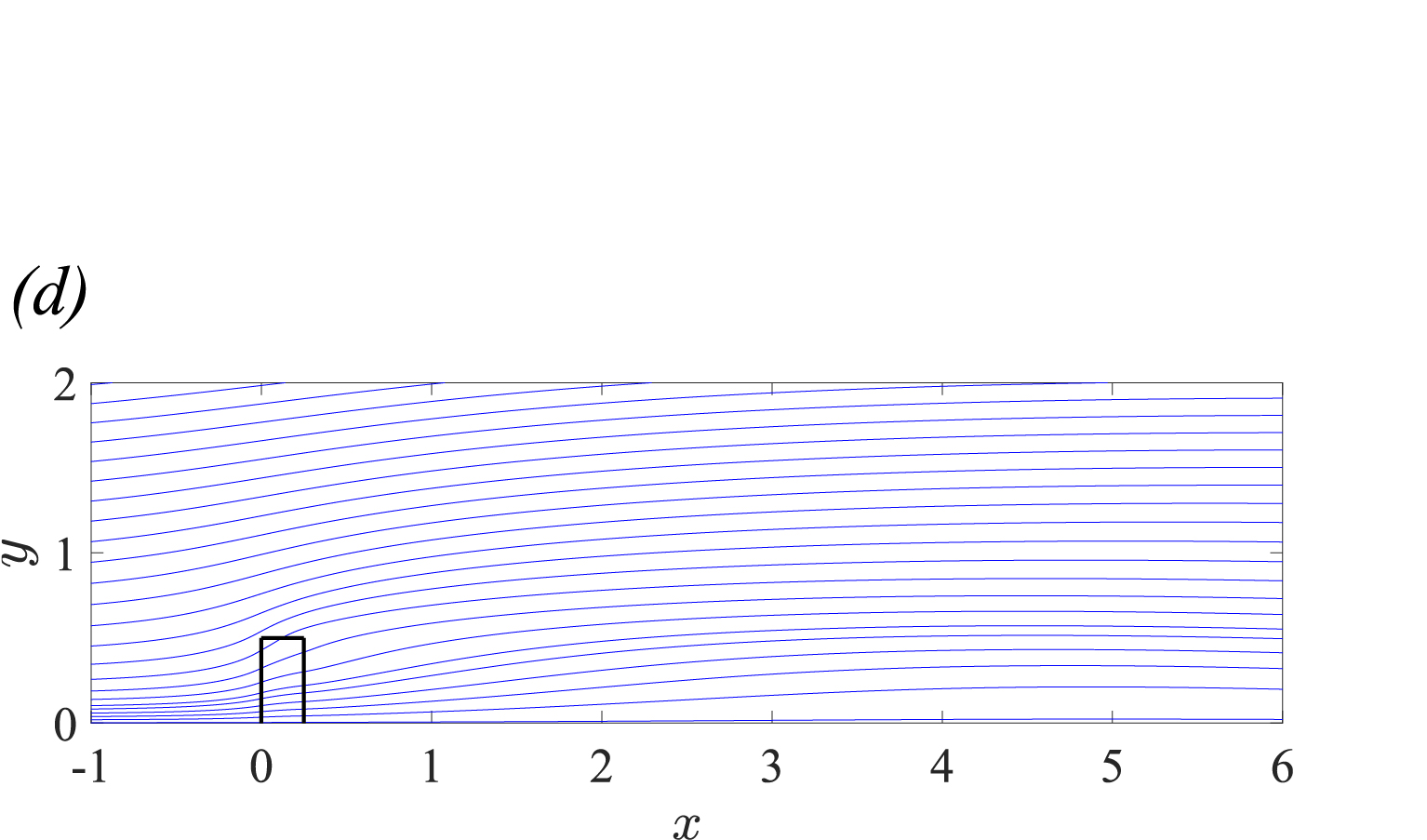}\\
\caption{Baseflow. Flow streamlines at $Da=1.1 \times 10^{-3}$, and (a) $Re=40$, (b) $Re=90$, (c) $Re=110$, (d) $Re=140$ (only half of the domain, i.e. $y\ge 0$, is shown).}
\label{fig:flowConf1}
\end{center}
\end{figure}
\FloatBarrier

\begin{figure}[h!]
\begin{center}
\includegraphics[trim=0mm 0mm 15mm 0mm,clip,width=.32\textwidth]{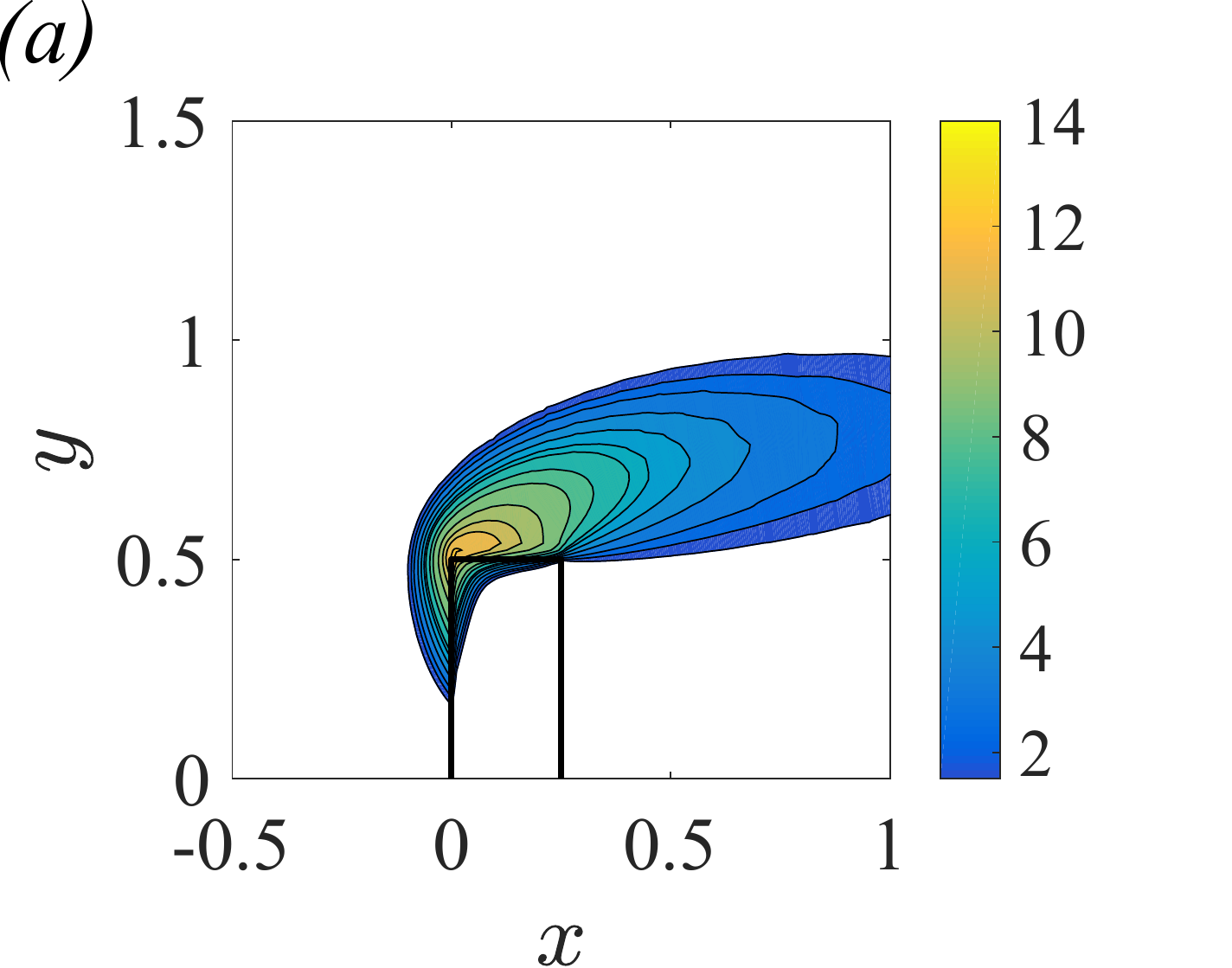}
\includegraphics[trim=0mm 0mm 15mm 0mm,clip,width=.32\textwidth]{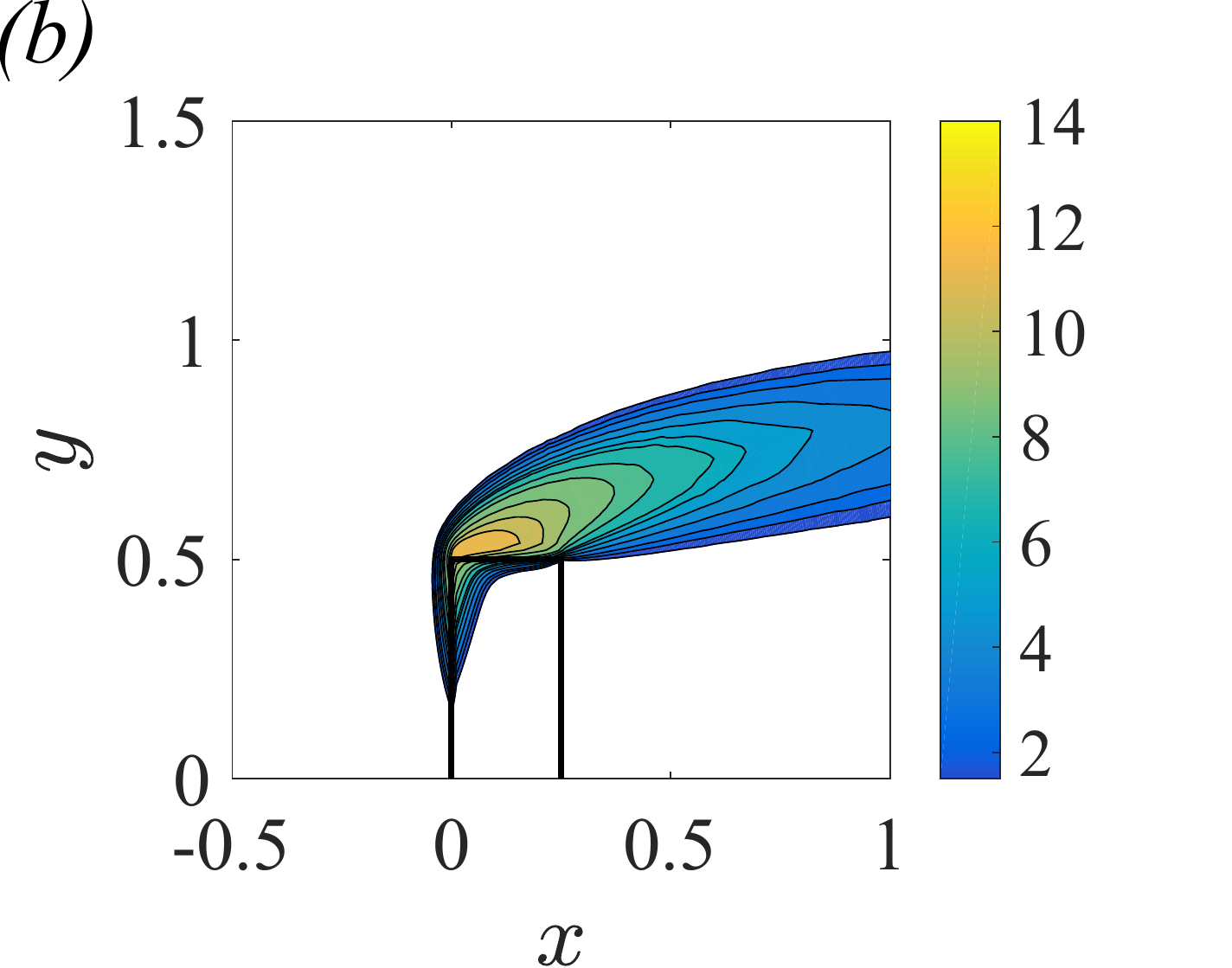}
\includegraphics[trim=0mm 0mm 15mm 0mm,clip,width=.32\textwidth]{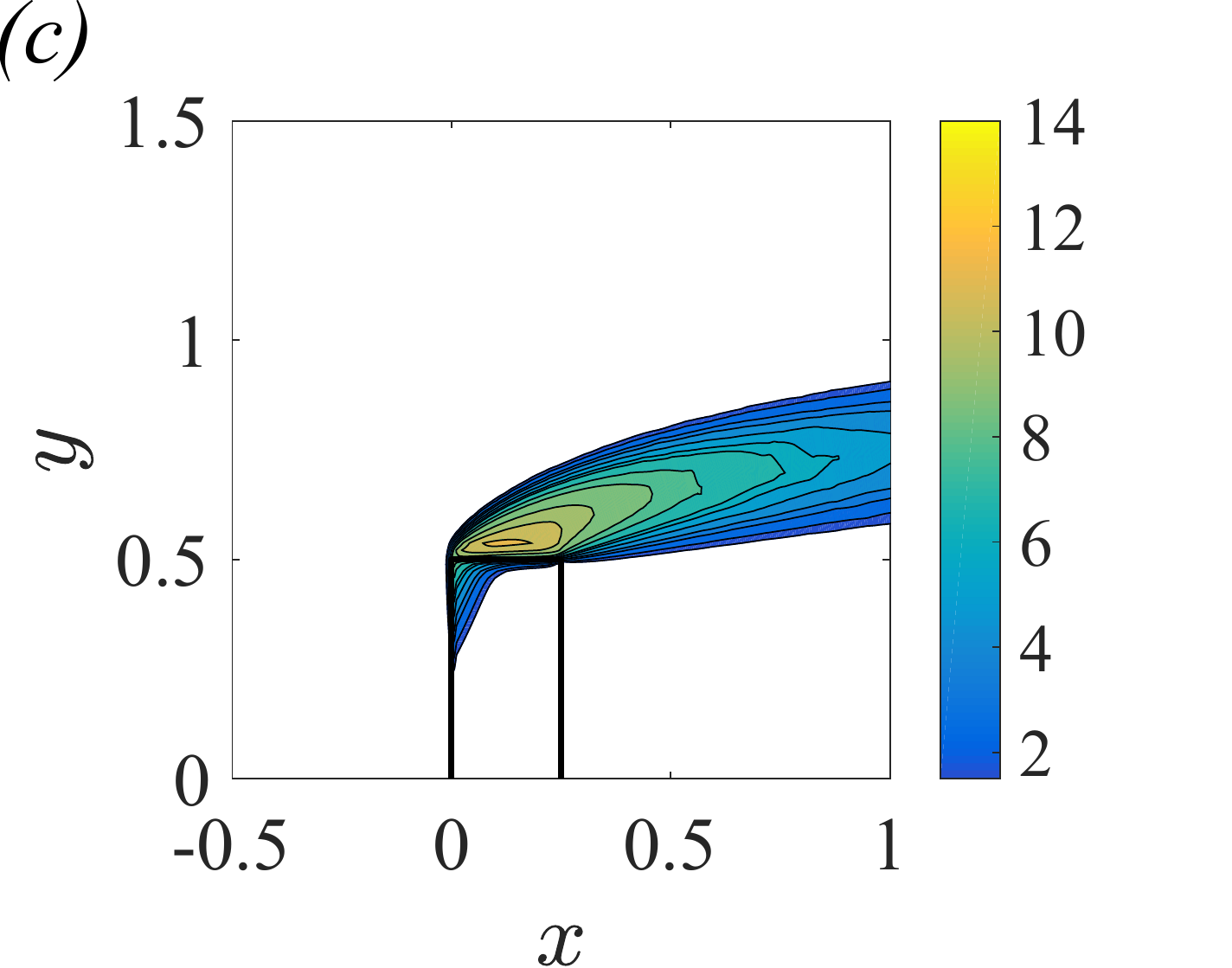}
\caption{Distribution of vorticity in the upper part of the body, at $Da=1.1 \times 10^{-3}$, and (a) $Re=40$, (b) $Re=65$, (c) $Re=110$.}
\label{fig:vort3}
\end{center}
\end{figure}

\begin{figure}[h!]
\begin{center}
\includegraphics[width=0.9\textwidth]{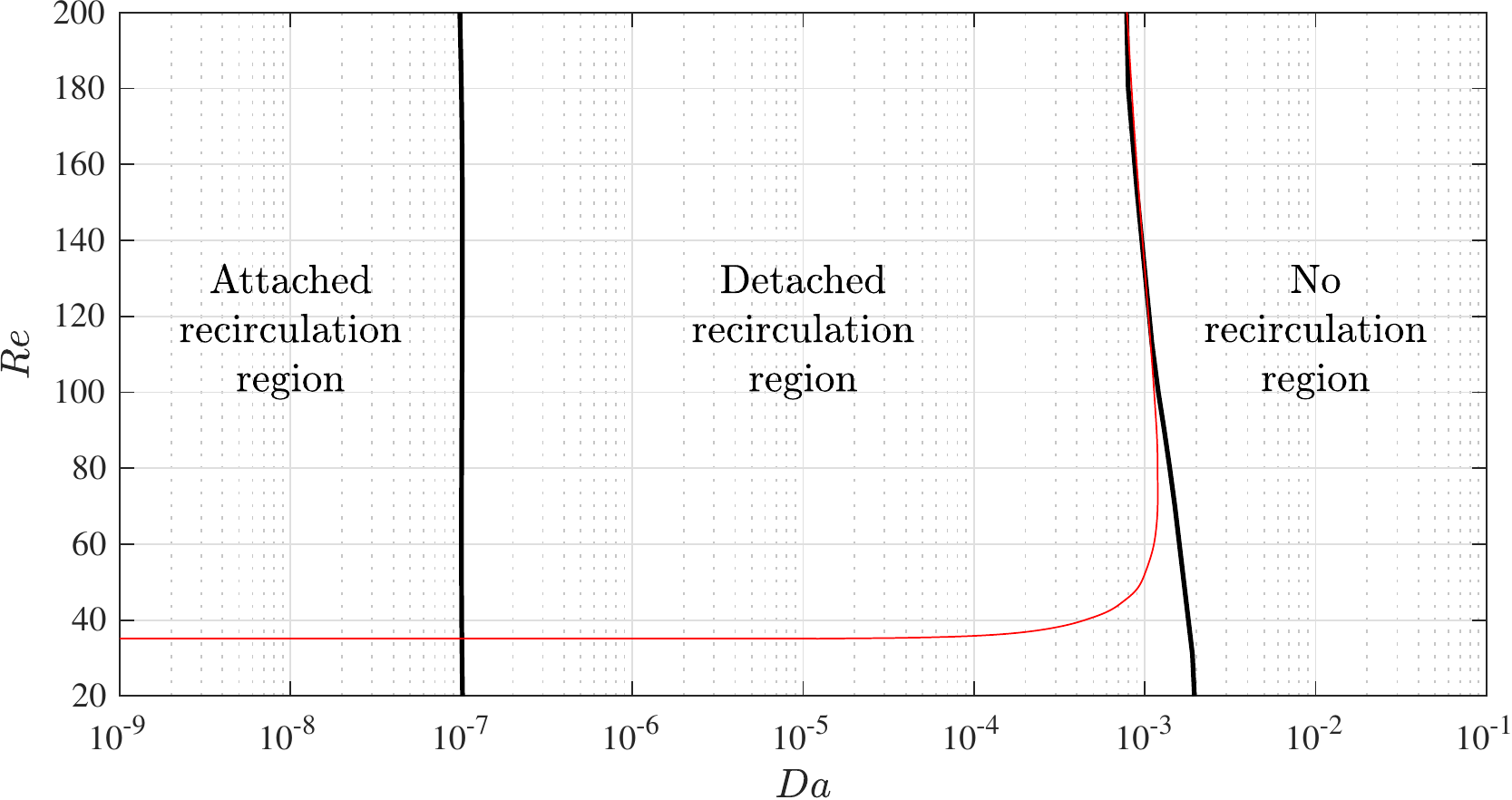}
\caption{Marginal stability curve (red line) in the $Re$-$Da$ plane, and the different flow patterns of the baseflow (delimited by the black lines).}
\label{fig:diagrammaBIF}
\end{center}
\end{figure}

\FloatBarrier
The set of transitions in the flow morphology described above are explored by varying both the flow Reynolds number and the Darcy number, and results are summarized in Fig. \ref{fig:diagrammaBIF}, where the type of wake flow, i.e. with attached, detached and no recirculation regions, is delimited by the black lines in the $Da-Re$ plane, for Reynolds numbers up to $Re=200$. It is possible to observe that the first critical Darcy value $Da_{cr1}$ is almost constant at $1\times10^{-7}$ and independent of the flow Reynolds number. On the other hand, the second critical Darcy number $Da_{cr2}$, that separates the cases with detached recirculation regions from cases without recirculation regions, is  slightly decreasing with the Reynolds number, reaching the value of $Da_{cr2}=8\times10^{-4}$ at $Re=200$.

The wake modifications due to the permeability of the porous medium directly affect the drag force $F_D$ on the body. When small values of $Da$ are considered, the drag coefficients $C_D$, here defined as $C_D=F_D/(0.5\rho U_{\infty }^2 t d)$, is very similar to the case of the solid cylinders. Referring to Fig. \ref{fig:CD}, the drag coefficient is $C_D\approx C_{D,solid}=2.18$ at $Re=20$ and it is $C_D\approx C_{D,solid}=1.88$ at $Re=30$. Increasing the Darcy number, the $C_D$ slightly decreases first and a significant reduction is successively visible for $Da>1\times10^{-3}$. In this range of $Da$, the drag coefficient follows the scaling $C_D \sim Re^{-1}Da^{-1}$, as reported in \cite{Cummins2017}.

\begin{figure}[h!]
\begin{center}
\includegraphics[width=0.7\textwidth]{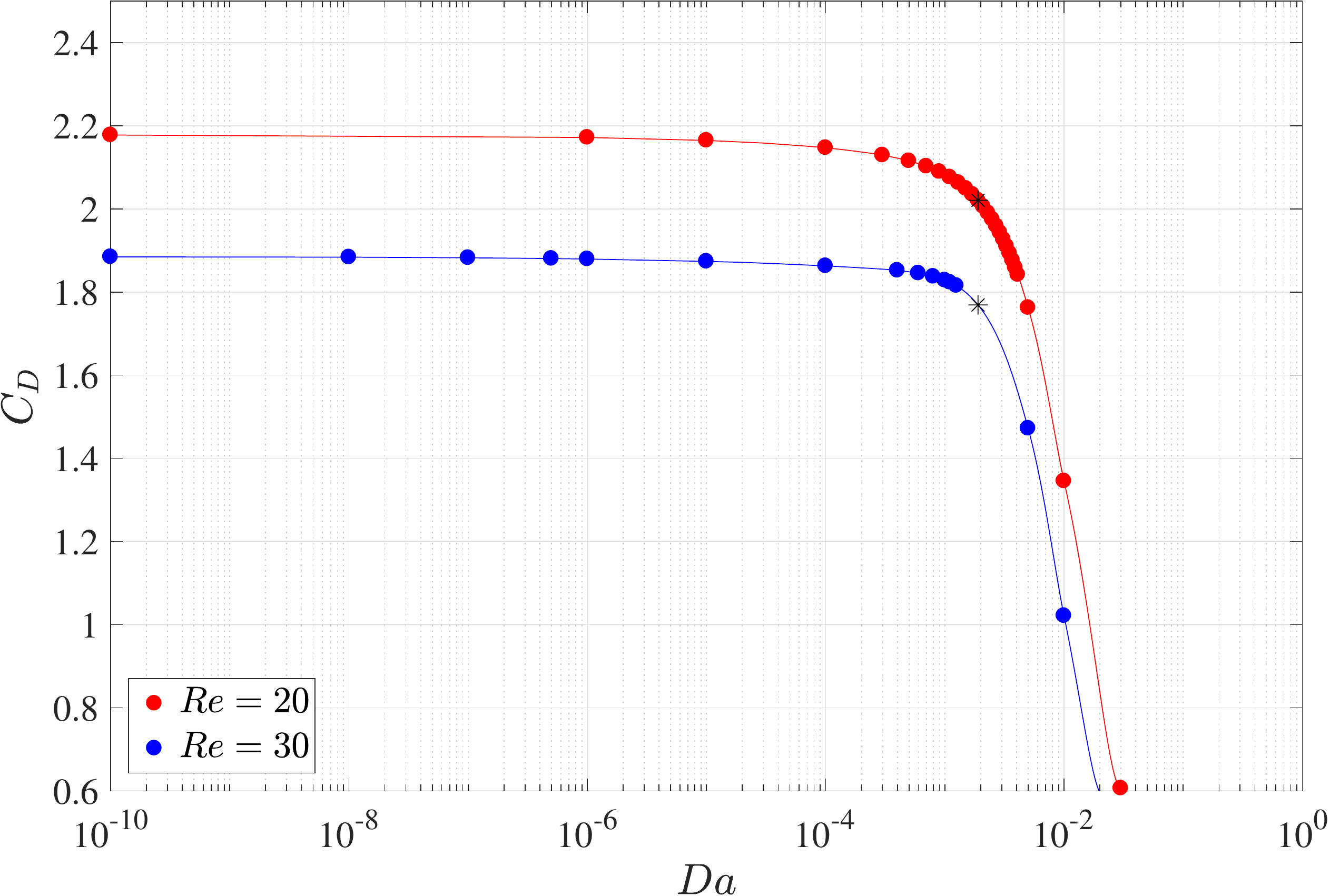}
\caption{Drag coefficient $C_{D}$ as a function of $Da$ for $Re=20$ (red line) and $Re=30$ (blue line).}
\label{fig:CD}
\end{center}
\end{figure}

\FloatBarrier
\subsubsection{Global stability analysis}
In this section the results of the stability analysis for the wake flow of the porous rectangular cylinder with $t/d=0.25$ are presented for Reynolds numbers up to Re=200 and varying the Darcy number. As shown in the previous section, the characteristics of the porous medium affect the behaviour of the wake flows and, consequently, a strong modification of the stability properties can be expected in comparison to the solid case. In particular, this latter case shows a Hopf bifurcation that drives the flow field from a symmetric solution, presented in the previous section, to a state which is periodic in time. This transition occurs for the solid case at a critical Reynolds number $Re_{cr} \approx 35$ and the resulting flow field is characterized by a nondimensional time frequency equal to $St={fd}/{U_{\infty}}\approx0.106$.

The results from the global stability analysis applied to the porous cases confirm that the nature of the instability is preserved for a wide range of values of the permeability. However, it is possible to identify configurations where the steady and symmetric solution remain stable for all the Reynolds numbers in the range considered here. This behaviour is shown in Fig. \ref{fig:diagrammaBIF}, which reports the neutral stability curve (red line), i.e. the curve that corresponds to the cases with null growth rate, $\lambda_0=0$, for different values of $Da$, together with the boundary curves that identify the different base flow configurations described in the previous section. It is clear that exists a threshold in permeability exists, $Da^{stab}_{cr}$, beyond which the occurrence of the Hopf bifurcation is suppressed. The value of the $Da^{stab}_{cr}$ depends on the Reynolds number and it reaches a maximum value of $1.2 \times 10^{-3}$ for $Re \simeq  80$, and it decreases as $Re$ is further increased. This behaviour is highlighted in Fig. \ref{fig:IsoDeltaL}, where the neutral curve is reported together with the iso-contours of $\Delta L$ in the $Re-Da$ plane. It results that the lower branch of the neutral curve follows the iso-contours of $\Delta L$ for a wide range of $Da$ up to $Da=Da^{stab}_{cr}$. For higher $Re$, the neutral curve crosses the iso-contours and enters in the area of the flow parameters where no recirculation bubbles is present in the base flow. 

\begin{figure}[h!]
\begin{center}
\includegraphics[width=.7\textwidth]{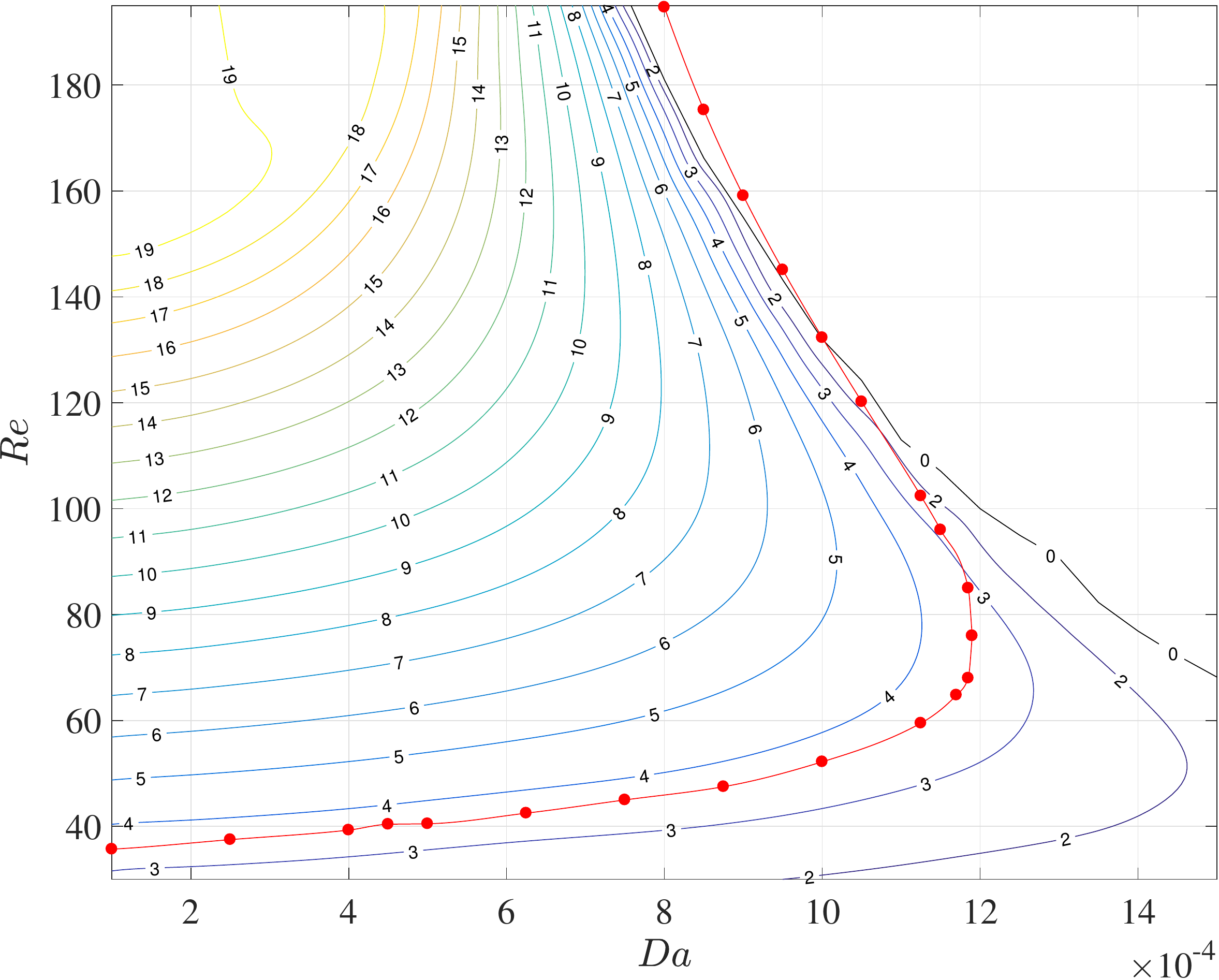}
\caption{Marginal stability curve (red line) and (a) iso-contours of the length of the recirculation bubble $\Delta L$ (colour lines), measured on the centerline.}
\label{fig:IsoDeltaL}
\end{center}
\end{figure}
\FloatBarrier
Thus, for particular couples of (Da,Re), the flow is globally unstable even if no recirculation regions are present, as for instance for $Da=8\times{10^{-4}}$ and Re=185. The possibility to have a global instability is in fact linked to the presence of a sufficiently strong wake defect and it is not directly   related to the presence of regions of counterflow (see for example \cite{MONKEWITZ88},  \cite{Biancofiore11} for details).
It is also interesting to observe from fig. \ref{fig:IsoDeltaL} that, for some fixed values of $Da$, e.g. $Da=10^{-3}$, the baseflow becomes first unstable and then recovers again a steady solution when the Reynolds number is further increased. 
The shapes of the global modes associated with the leading global eigenvalues are affected by the characteristics of the base flow. In particular, the downstream displacement of the recirculation regions suggests that also the perturbations originate in a region which moves progressively downstream as Re is increased. This behaviour is visible in Fig. \ref{fig:direct}, where the marginal global modes for $Re=52$ and $Re=132$ at $Da=10^{-3}$ are reported. 

From the literature it is  known that the wavemaker of the vortex shedding instability is localised in the recirculation region past a solid bluff body (see, for example, \cite{Meliga09}). Consequently, it is expected that the wavemaker region for a porous body follows the position of the recirculation region or, in  general, the position of the maximum wake velocity defect. Following \cite{Giannetti07}, the wavemaker can be identified evaluating the inner product between direct mode and adjoint global mode. As an example, the adjoint leading modes are reported in Fig. \ref{fig:adjoint}, for the same cases of the direct modes shown in Fig. \ref{fig:direct}.
The shape of the structural sensitivity is finally reported in Fig. \ref{fig:structsens}. \\

\begin{figure}[h!]
\begin{center}
\includegraphics[trim=0mm 145mm 0mm 130mm,clip,width=.8\textwidth]{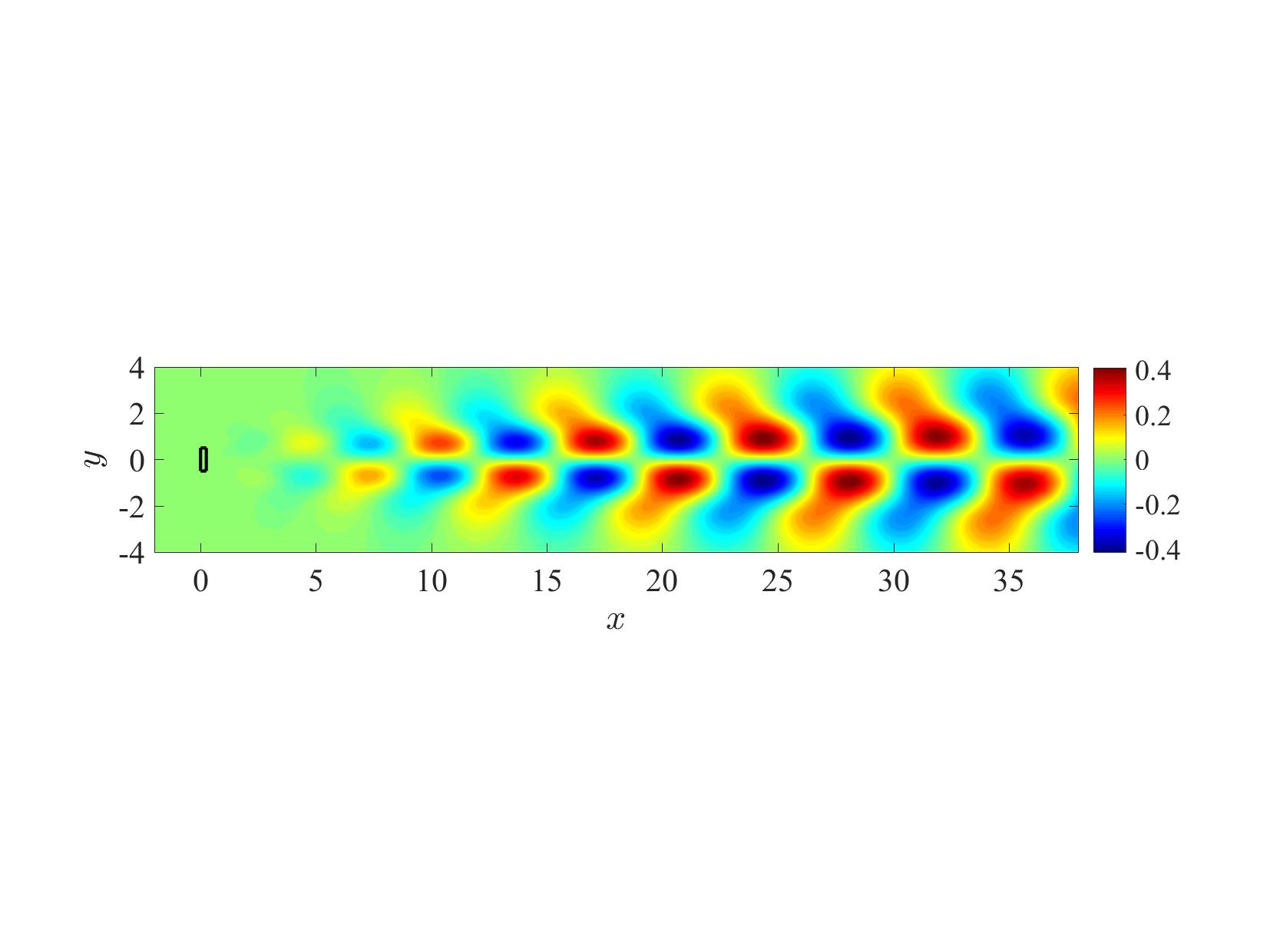}\\
\includegraphics[trim=0mm 145mm 0mm 130mm,clip,width=.8\textwidth]{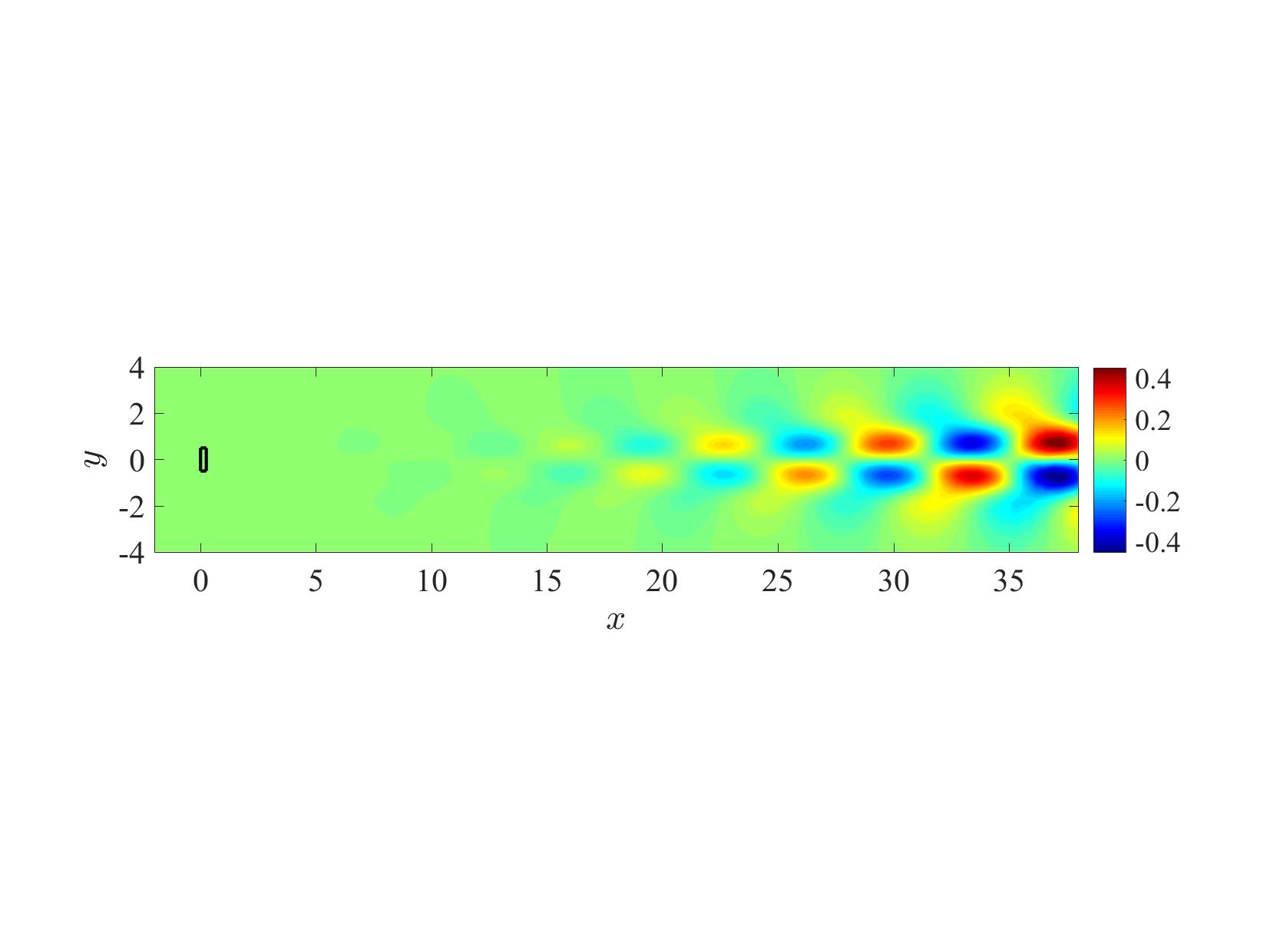}\\
\caption{Real part of the streamwise component of the direct eigenvector at $Da=10^{-3}$ and (a) $Re=52$, (b) $Re=132$, both on the marginal stability curve.}
\label{fig:direct}
\end{center}
\end{figure}

\begin{figure}[h!]
\begin{center}
\includegraphics[trim=0mm 90mm 0mm 90mm,clip,width=.495\textwidth]{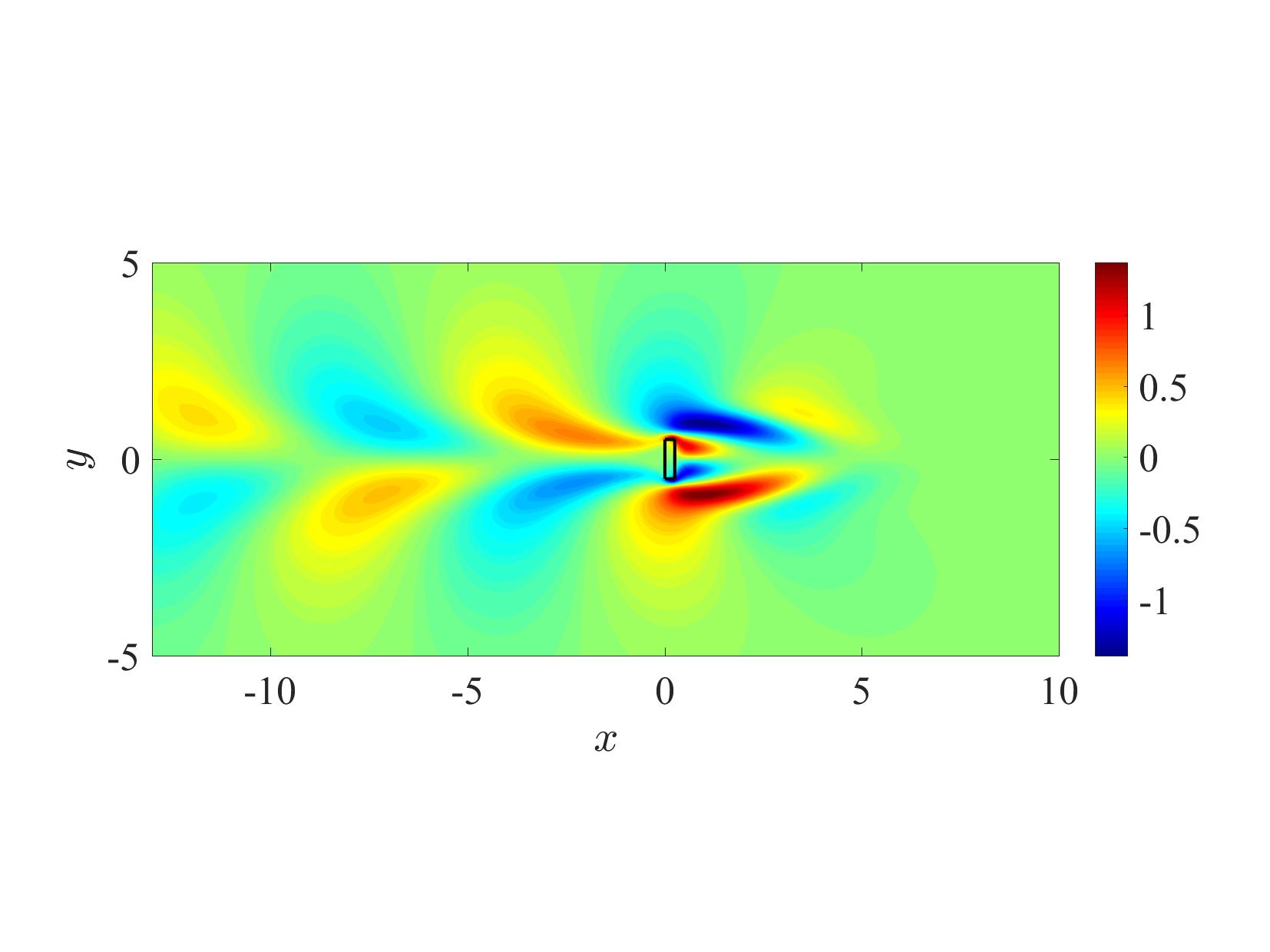}
\includegraphics[trim=0mm 90mm 0mm 80mm,clip,width=.495\textwidth]{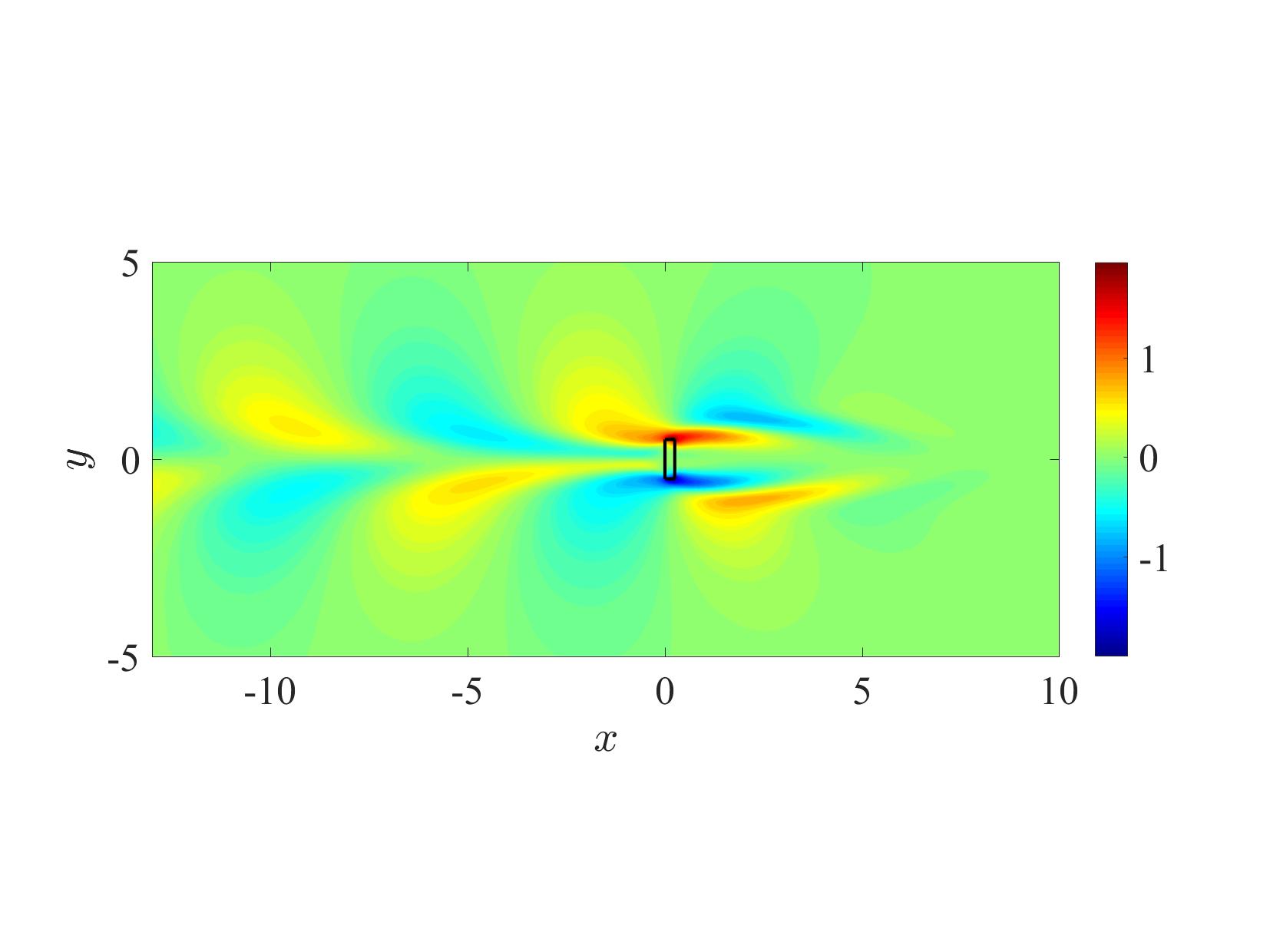}
\caption{Real part of the streamwise component of the adjoint eigenvector at $Da=10^{-3}$ and (a) $Re=52$, (b)$Re=132$.}
\label{fig:adjoint}
\end{center}
\end{figure}

\begin{figure}[h!]
\begin{center}
\includegraphics[trim=0mm 130mm 0mm 90mm,clip,width=.495\textwidth]{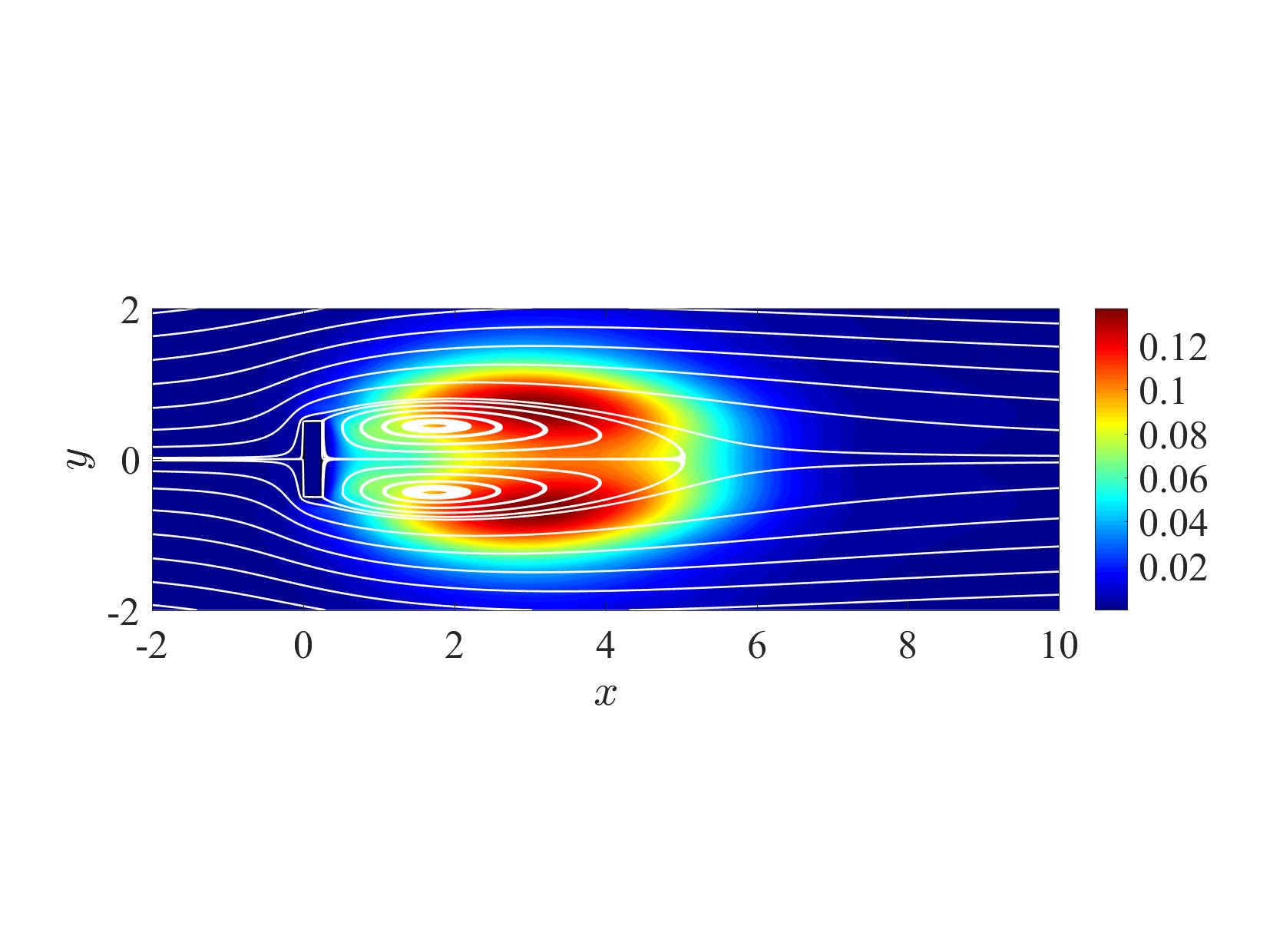}
\includegraphics[trim=0mm 130mm 0mm 90mm,clip,width=.495\textwidth]{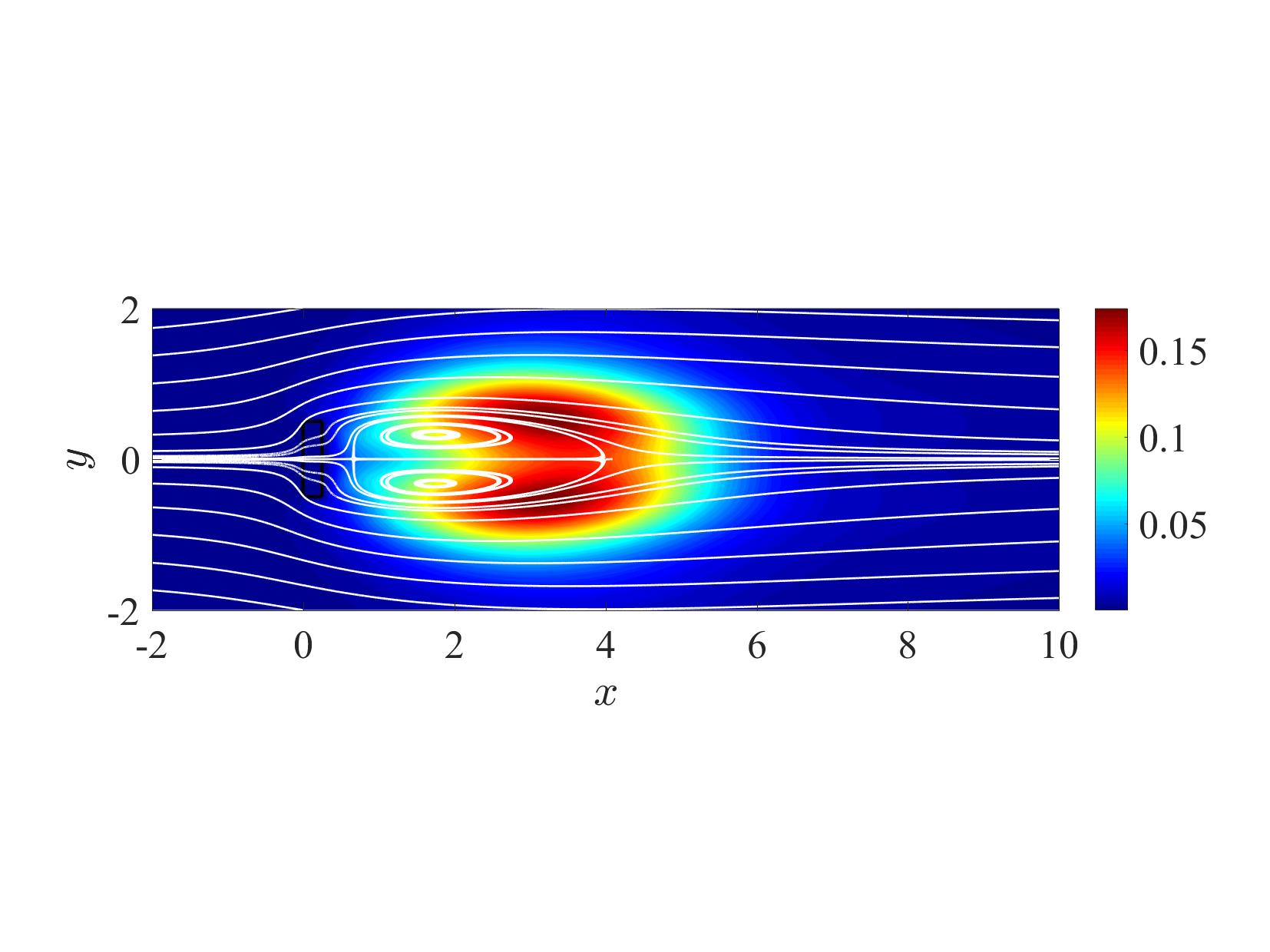}\\
\includegraphics[trim=0mm 100mm 0mm 90mm,clip,width=.495\textwidth]{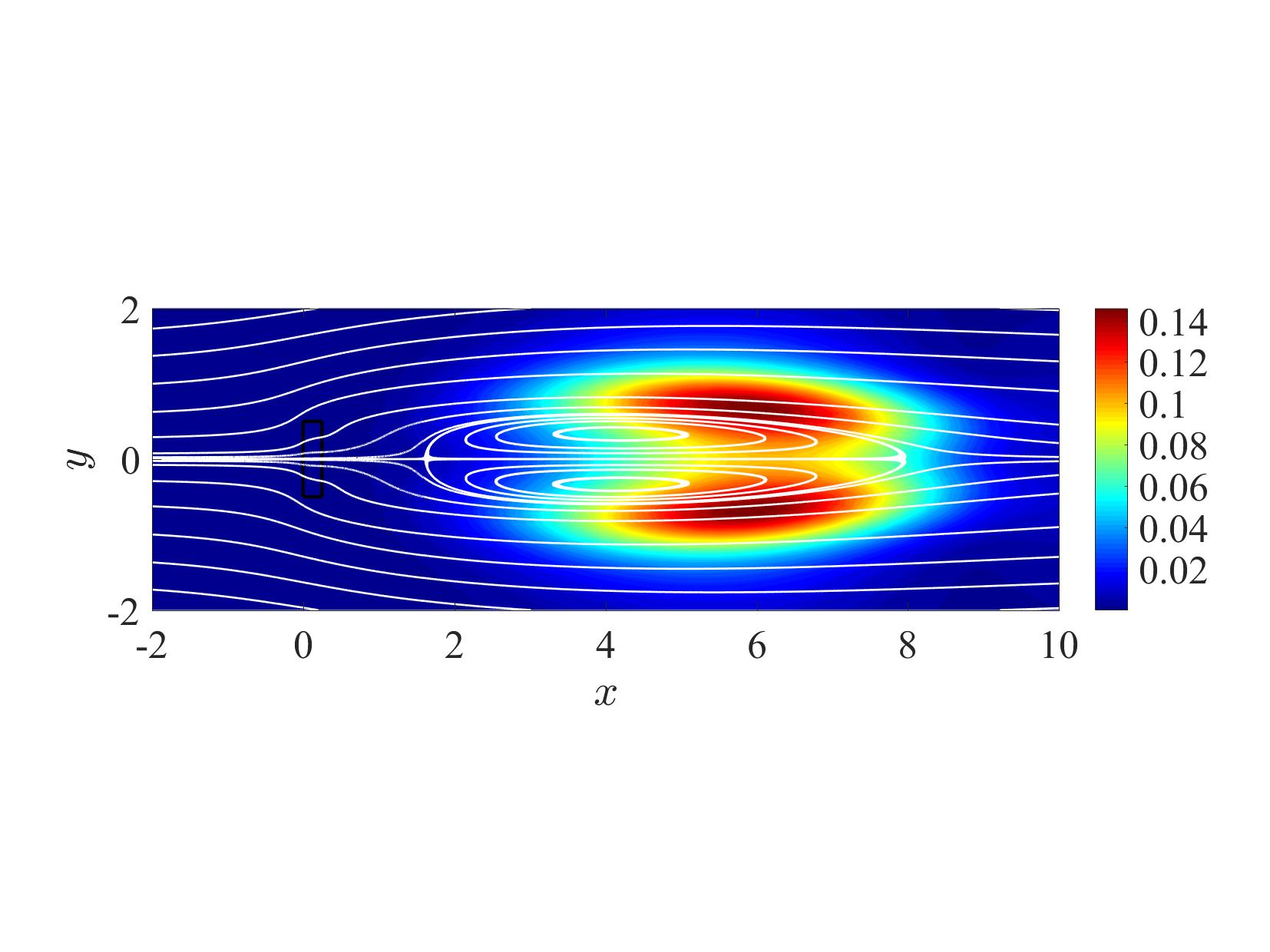}
\includegraphics[trim=0mm 100mm 0mm 90mm,clip,width=.495\textwidth]{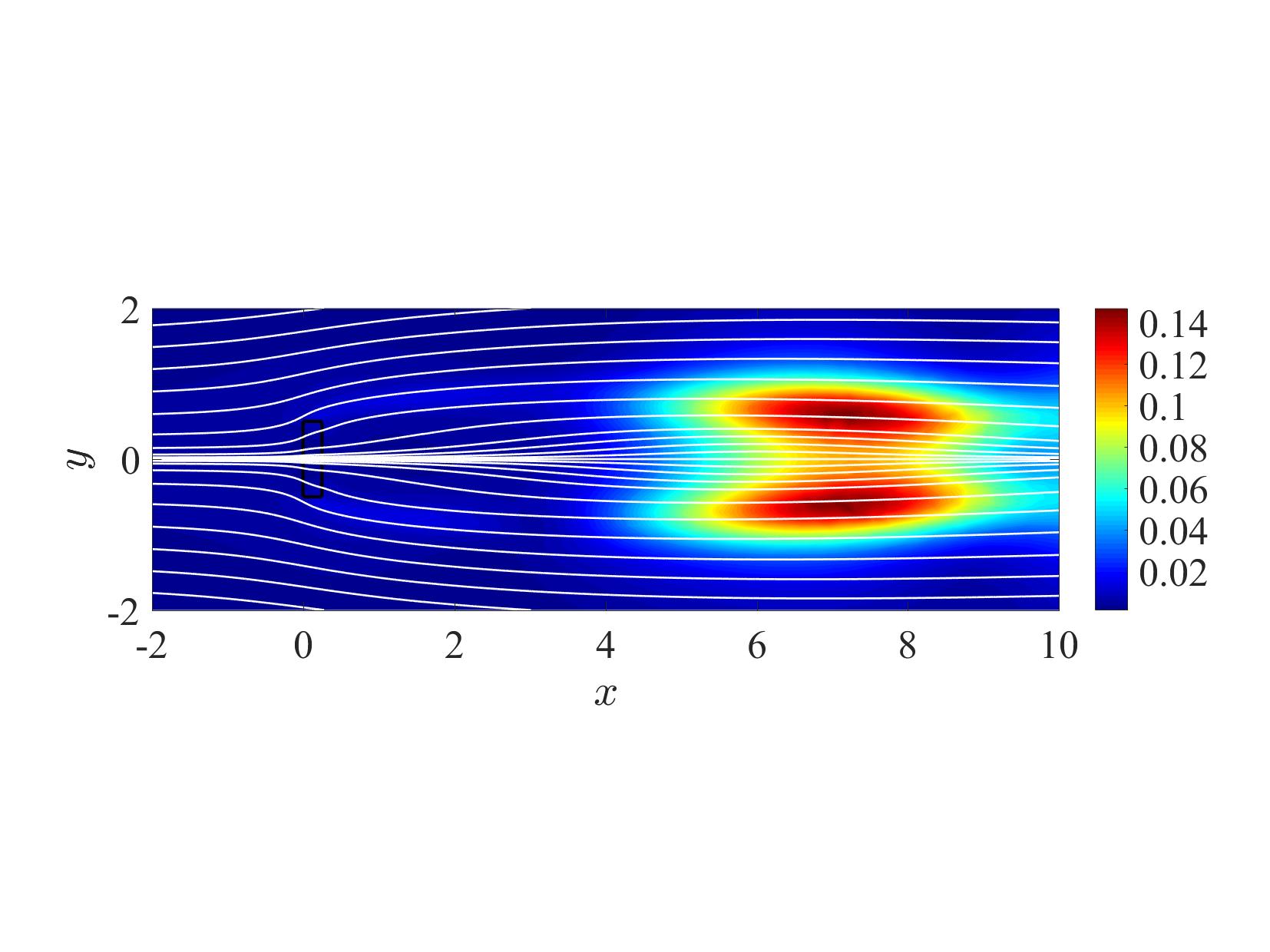}\\
\caption{Sensitivity to structural perturbations of the bifurcating global mode at (a) $Re=45$, $Da=10^{-10}$, (b) $Re=54.5$, $Da=9 \times 10^{-4}$, (c) $Re=100$, $Da=9 \times 10^{-4}$, (d) $Re=160$, $Da=9 \times 10^{-4}$.}
\label{fig:structsens}
\end{center}
\end{figure}
\FloatBarrier
When low values of $Da$ are considered, the wavemaker is located close to the body similarly to the solid case (Fig. \ref{fig:structsens}a). However, when $Re$ and $Da$ are increased, the wavemaker moves downstream together with the recirculation regions (Fig. \ref{fig:structsens}b,c). Finally, even if the recirculation bubbles are not present anymore, the wavemaker still persists in the velocity deficit region (Fig. \ref{fig:structsens}d). 
The results confirm that, even without a recirculation region, a sufficiently strong wake defect can sustain an unsteady global instability. 

In order to gain more insight on this aspect  the spatio-temporal stability properties of the flow are given in the next section.

\subsubsection{Spatio-temporal stability analysis}
\begin{figure}[h!]
\begin{center}
\includegraphics[width=.49\textwidth]{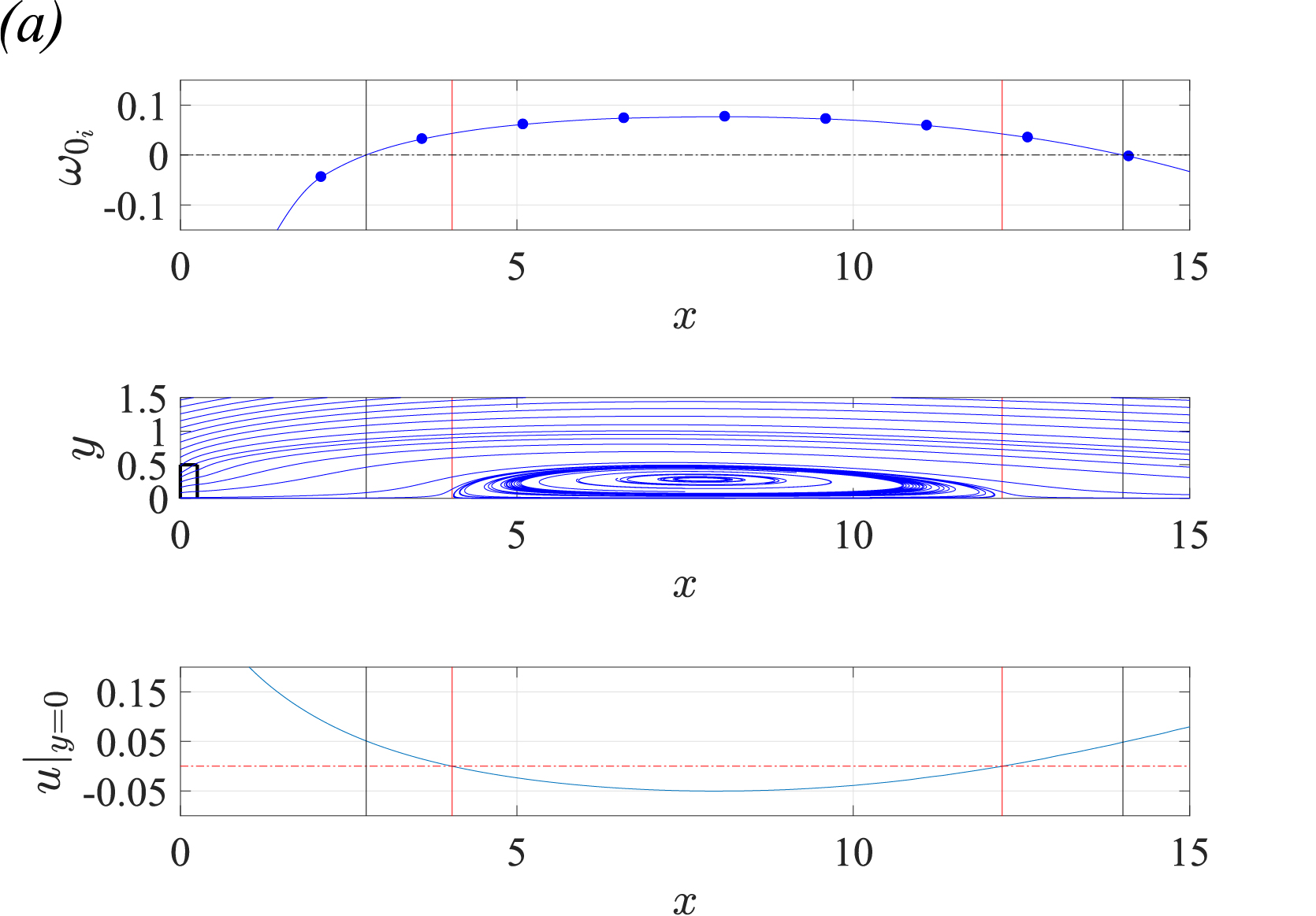}
\includegraphics[width=.49\textwidth]{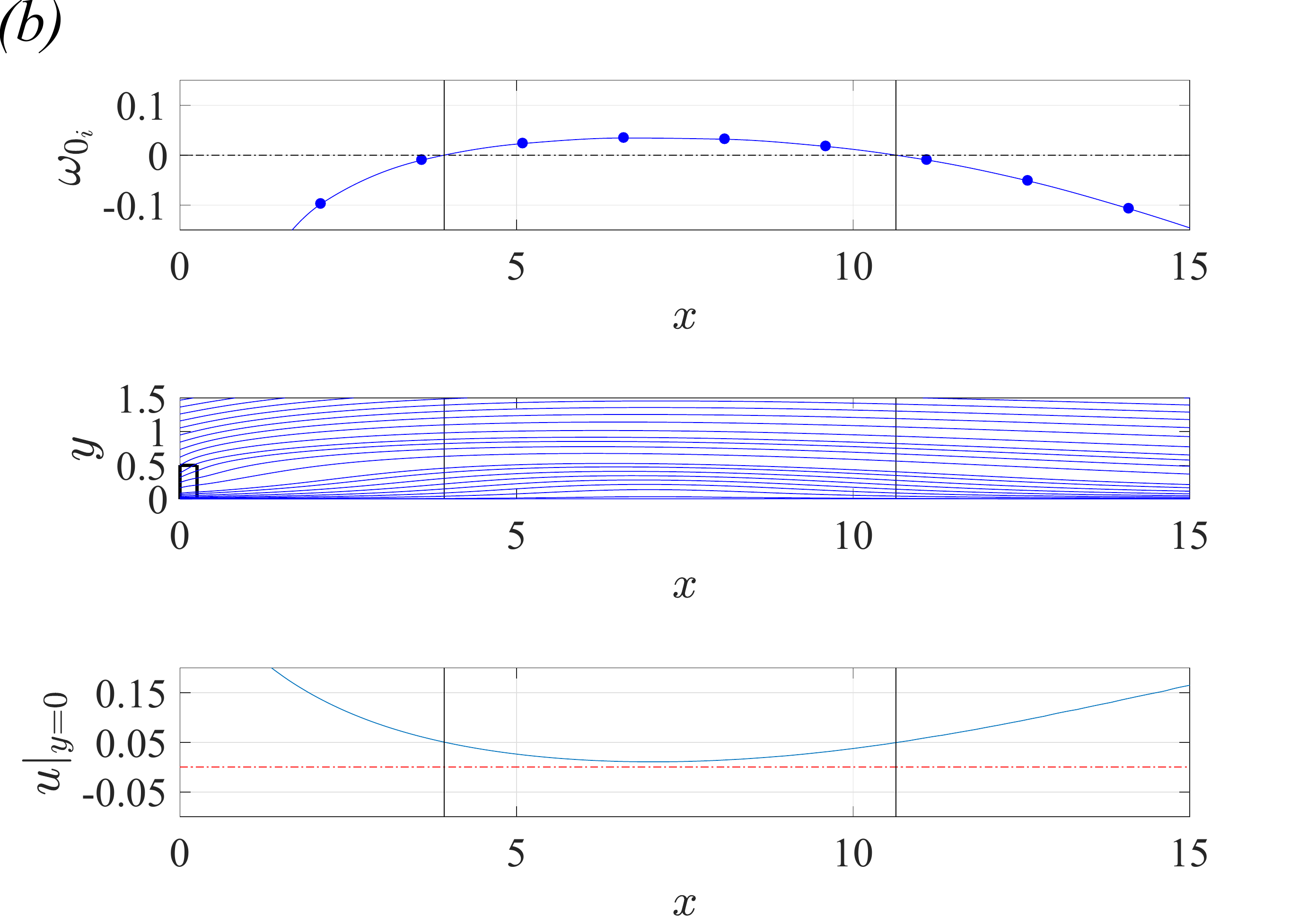}
\includegraphics[width=.49\textwidth]{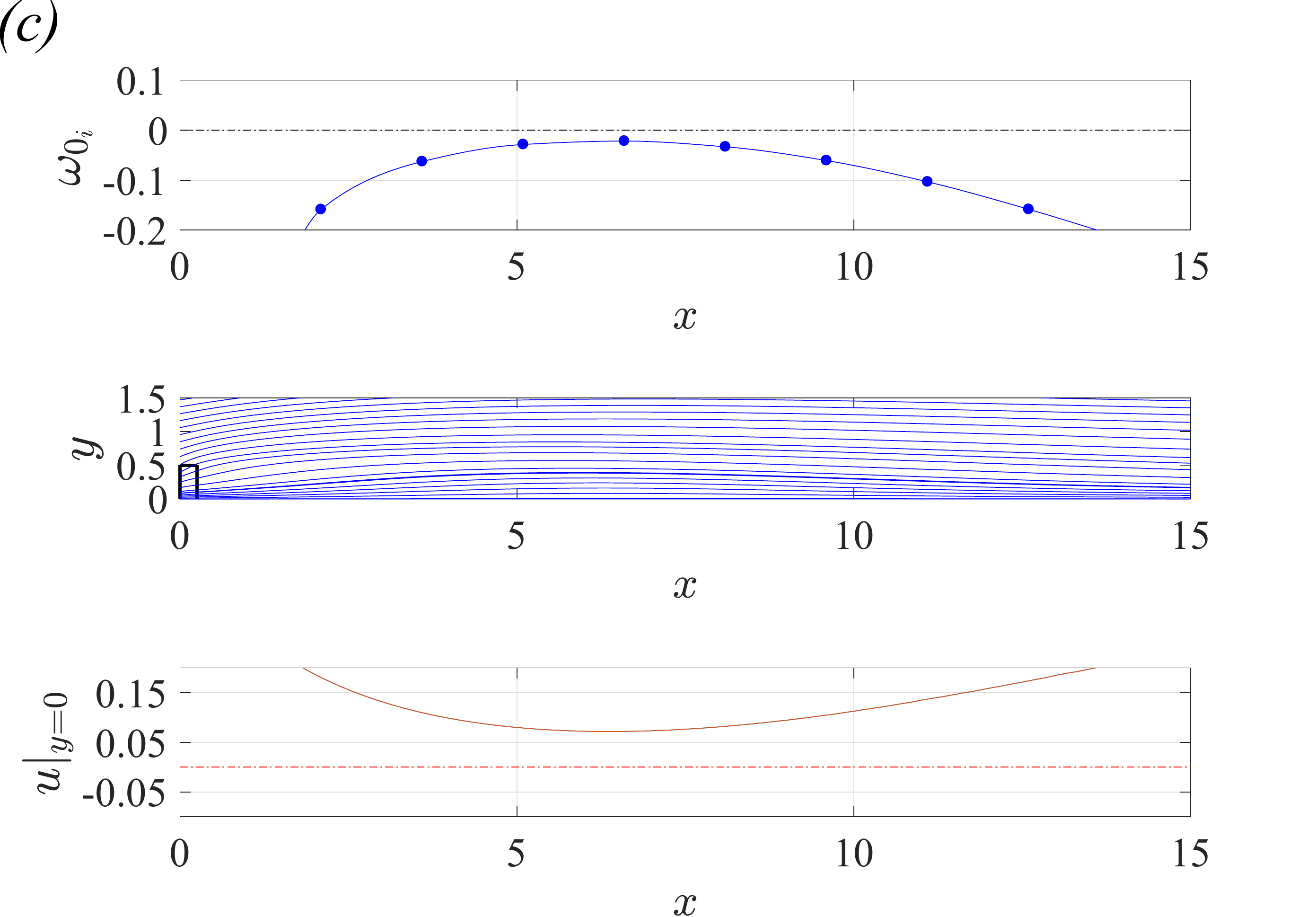}
\includegraphics[width=.46\textwidth]{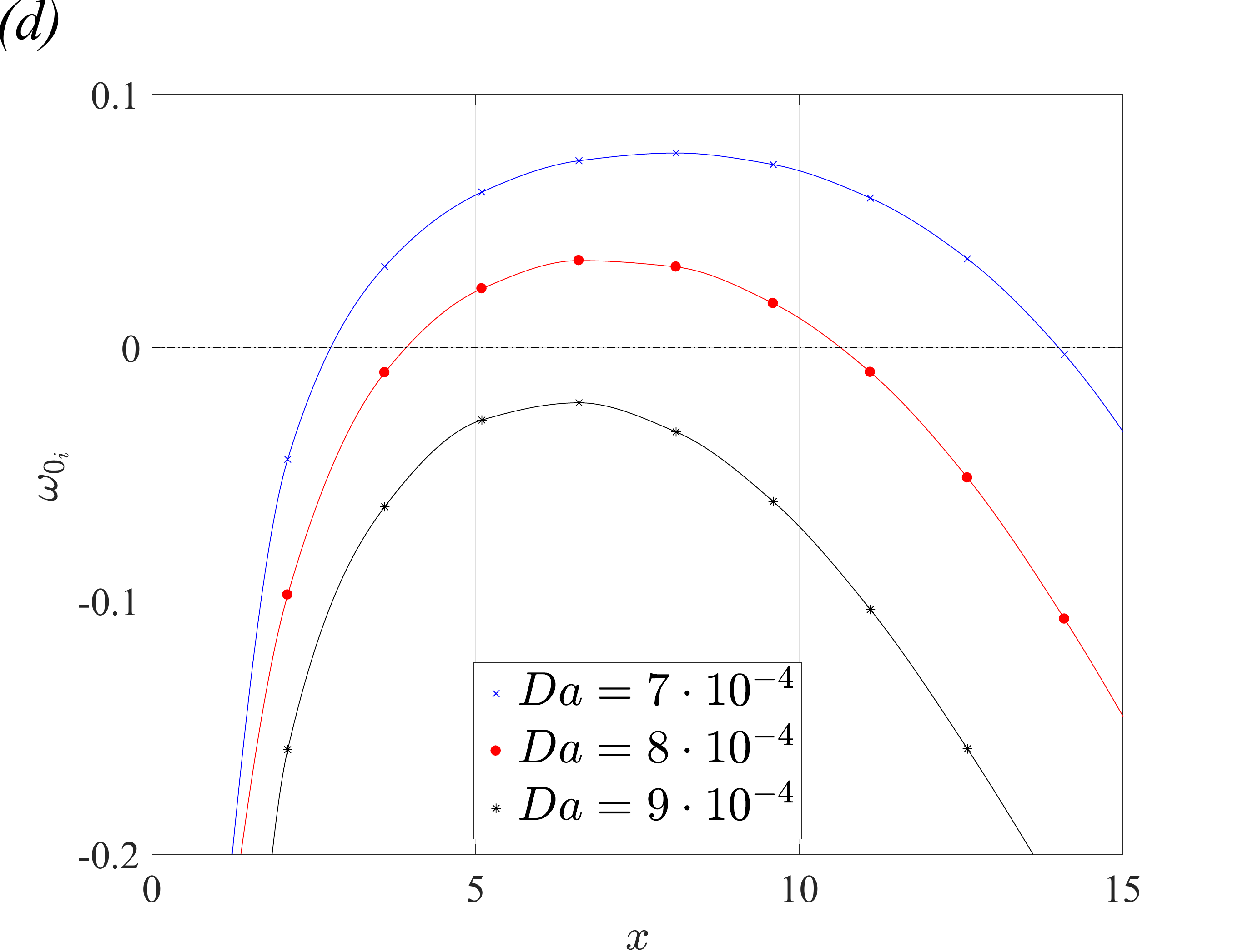}
\caption{Local stability properties of the wake behind a porous disk with $t/d = 0.25$, $Re=185$ and increasing $Da$. From top down, the absolute growth rate $\omega_{0i}$ at different streamwise locations, the baseflow streamlines, and the streamwise component of the velocity evaluated at the centerline, for (a) $Da=7 \times 10^{-4}$, (b) $Da=8 \times 10^{-4}$,(c) $Da=9 \times 10^{-4}$. (d) The absolute growth rate $\omega_{0i}$ for the three cases reported.}
\label{fig:stabassRe185}
\end{center}
\end{figure}

In this section, the local stability properties of the wake past porous cylinders are analyzed in the framework of spatio-temporal stability analysis. In particular, the representative case of Reynolds number of $Re=185$ and aspect ratio $t/d=0.25$ is here discussed for three values of permeability, one globally unstable with a recirculation region, one globally unstable but without recirculation and one which is globally stable. The objective is to investigate the region of absolute instability in the three considered wakes so as to provide a further viewpoint so as to explain the behavior observed by global stability analysis described in the previous section. Specifically, at $Da= 7 \times 10^{-4}$, the base flow is globally unstable and it is characterized by a large absolute unstable region, which includes the recirculation region in the wake (Fig. \ref{fig:stabassRe185}a). The extension of the absolute region corresponds indeed to the locations where the streamwise velocity at $y=0$ is less then $0.05U_{\infty}$, according to the results of \cite{MONKEWITZ88}. As anticipated, counterflow is not necessary for the wake profile to be absolutely unstable. As a consequence, there exists a range of $Da$ such that the recirculation bubble is not present, but the a global unstable mode is supported by a sufficiently elongated region of absolute instability provided that the wake deficit is stronger than approximately $5\%$. This scenario is confirmed in Fig. \ref{fig:stabassRe185}b, where the Darcy number is set to $Da= 8 \times 10^{-4}$.
Finally, further increasing the permeability, the wake deficit is recovered and the absolute region reduces or, eventually, disappears and, as a results, all the wake profiles becomes convectively unstable. In this case the flow becomes globally stable, as in Fig. \ref{fig:stabassRe185}c for $Da=9 \times 10^{-4}$. 
The results in terms of absolute growth rate are then summarized in Fig. \ref{fig:stabassRe185}d, where the stabilizing effect of the permeability, $k$, on the absolute unstable regions can be observed. 

Summarising, the results of this section show the link between the absolute instability region and the global instability of the considered wakes. It is shown that the region of absolute instability moves downstream together with the recirculation region or, in general, with the region where the wake velocity defect is concentrated. Moreover, it is clear by this analysis that global instability is related to the velocity defect more than to recirculation regions, providing further quantitative support to what observed by global stability analysis for those unstable configurations where recirculations are absent. 

\subsection{Effect of the porosity and of the aspect ratios on the stability of porous rectangular cylinders.}
The previous section shows the results for configurations at different values of the permeability $Da$ but with fixed porosity $\phi=0.65$ and thickness-to-height ratio $t/d=0.25$. In this section, the effect of these two parameters on the flow characteristics is investigated. \\
\indent Firstly, the effect of the porosity on the stability characteristics is investigated for the case with $t/d=0.25$. The marginal stability curves and the region of the parameter space where the baseflow has not recirculation regions, i.e. $\Delta L=0$, have been evaluated for porosities of $\phi=0.80$ and $\phi=0.95$. The results, reported in Fig. \ref{fig:poro} together with the ones obtained for $\phi=0.65$, show that the porosity weakly affects the stability properties and the main discrepancies are concentred in the upper branches of the neutral curves. In general, the qualitative behaviour is preserved and, moreover, in all the cases the iso-contours of $\Delta L=0$ cross the neutral curves, identifying regions in the (Re,Da) space where the self-sustained oscillations are present even in absence of recirculation regions.
\begin{figure}[t!]
\begin{center}
\includegraphics[width=.55\textwidth]{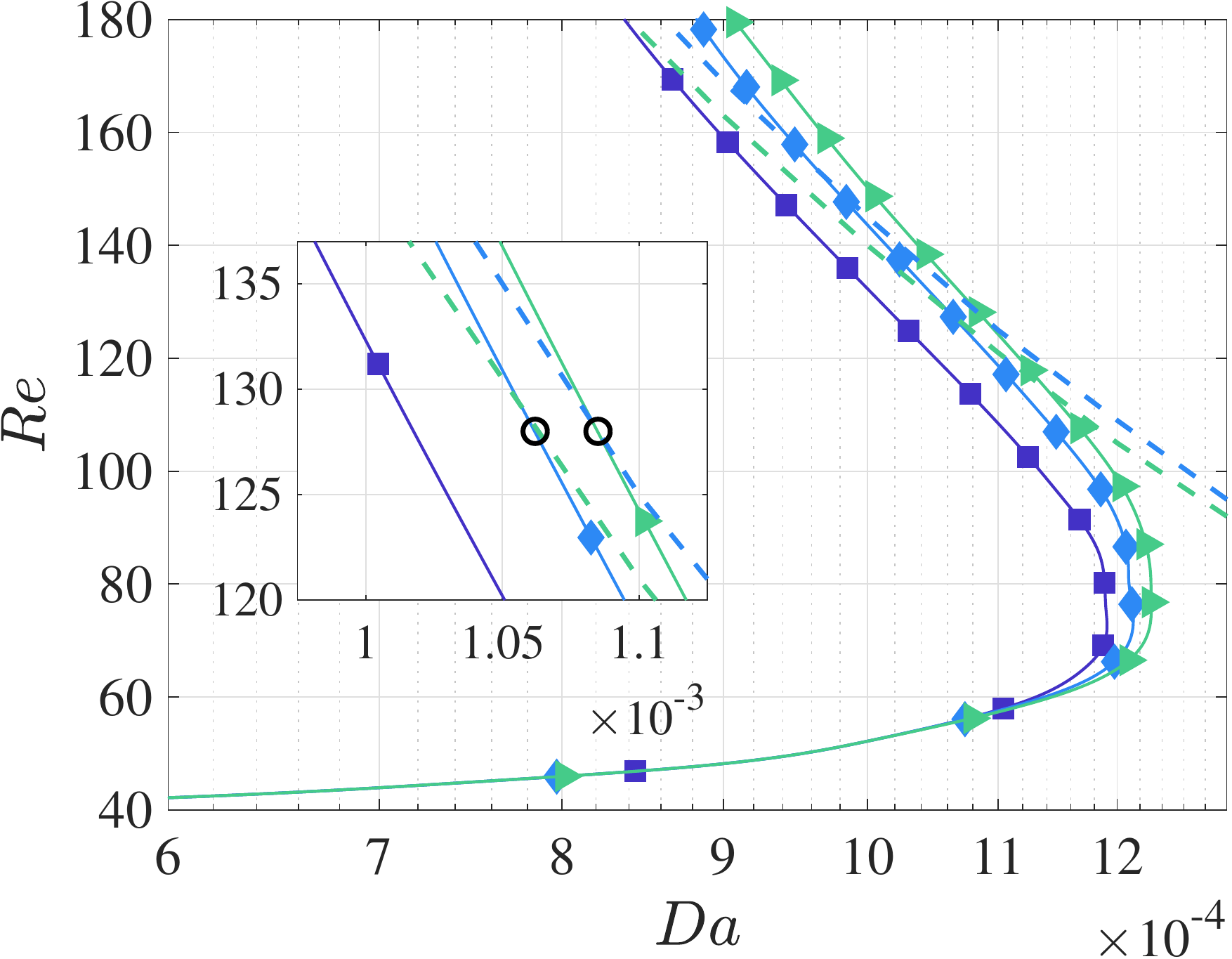}
\caption{Effect of the porosity on the bifurcation diagram and on the isocontour $\Delta L=0$, for $\phi=0.65$ $\blacksquare$, $\phi=0.8$ $\blacklozenge$ and $\phi=0.95$ $\blacktriangleright$; isocontours of $\Delta L=0$ are reported as dashed lines fo the two cases $\phi=0.8$ and $\phi=0.95$, highlighting with a black circle its intersection with the neutral stability curve (see the zoomed view at the center of the figure).}
\label{fig:poro}
\end{center}
\end{figure}

\indent Finally, the effect of the thickness-to-height ratio $t/d$ of the cylinder on the stability properties is studied for the case with porosity equal to $\phi=0.65$ and different values of $Da$. In particular, the $t/d$
is here varied from 0.01, i.e. thin plate configuration, to 1, i.e. square cylinder configuration. 
The results in terms of neutral stability curves, reported in Fig. \ref{fig:spessore}a normalizing the critical Reyonlds number with the corresponding values for the non-porous cases, show that the qualitative behaviour remains the same for all the cases, although the curves are shifted along the $Da-$axis. In particular, it is possible to observe that the variation of $t/d$ mainly modifies the contribution of the Darcy terms in the equations (\ref{DNSfinal2}) and, thus, the effect on the stability curve is expected to be linear with $t/d$. This speculation is indeed confirmed in Fig. \ref{fig:spessore}b, where the neutral stability curves are reported using the Darcy number based on the body's cross-section, i.e. $Da^*=Da \cdot (t/d)^{-1}=k/(td)$. {\lorenzo{All the curves roughly collapse onto the same curve, as clearly visibile in Fig. \ref{fig:spessore}b, especially for the cylinders with $t/d<0.5$. This new definition of the non-dimesional permeability $Da^*$ allows, finally, to identify a sharp threshold ${Da_T}^*=5 \times 10^{-3}$ (Fig. \ref{fig:spessore}b) valid for all the considered geometries beyond which the baseflow is always linearly stable and, then, the occurrence of time periodic wake solutions are unconditionally prevented in the parameter space here considered. }}

\begin{figure}[h!]
\begin{center}
\includegraphics[width=1\textwidth]{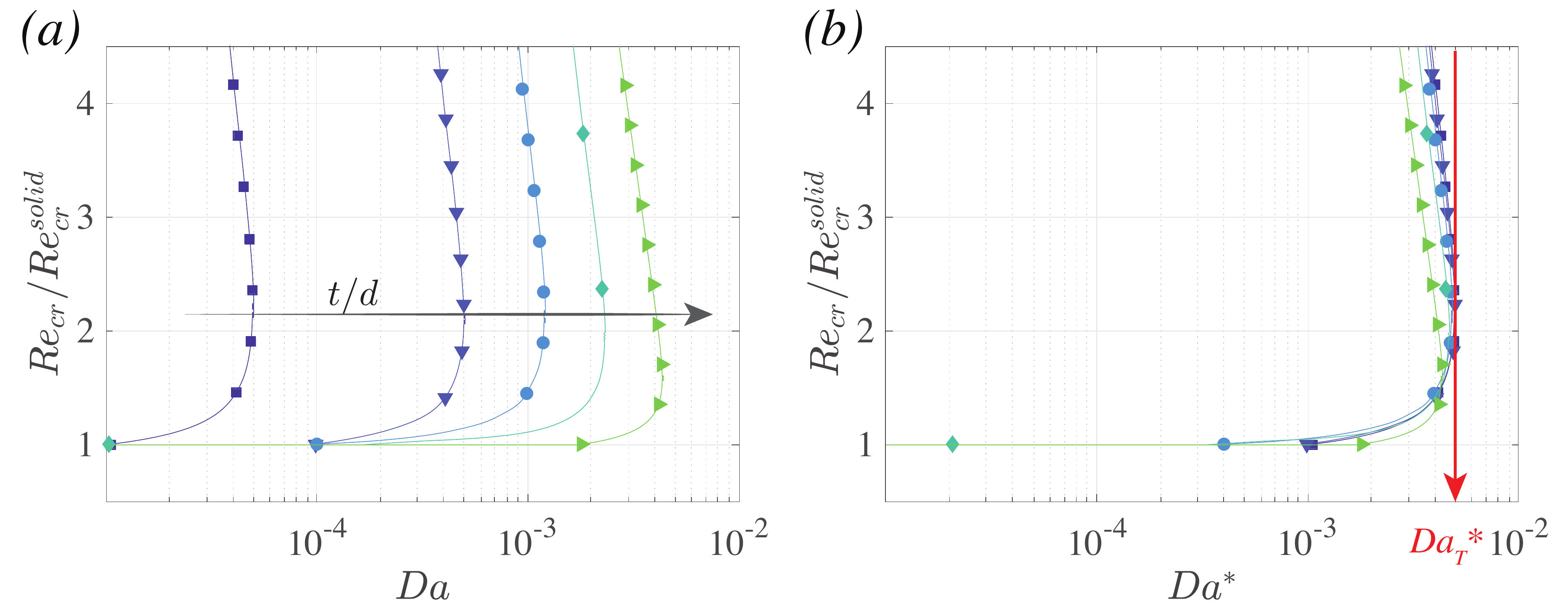}
\caption{Effect of the thickness-to-height ratio $t/d$ on the bifurcation diagram. (a) Neutral stability curves in the $Da-Re_{cr}$ plane and (b) using a modified $Da^*=Da \cdot (t/d)^{-1}=k/(td)$. The considered cases are: $t/d=0.01$ $\blacksquare$, $t/d=0.1$ $\blacktriangledown$,  $t/d=0.25$ $\bullet$, $t/d=0.5$ $\blacklozenge$, $t/d=1.0$ $\blacktriangleright$.}
\label{fig:spessore}
\end{center}
\end{figure}

\FloatBarrier
\section{Conclusions}\label{sec:conclusions}
\label{Conc}
In this work the characteristics and the stability properties of the steady flow around porous rectangular cylinders  at low-moderate Reynolds numbers  have been investigated. 
The problem has been tackled numerically using a mathematical model for the flow inside the porous medium that is based on the volume averaged Navier Stokes equations.
The resulting formulation, which takes into account both the viscous and inertial terms inside the porous medium, has been validated against direct numerical simulations documented in the literature. Once validated, the resulting numerical tools have been applied to study the flow past rectangular cylinders by systematically varying their aspect ratio, the permeability (by varying the Darcy number), the porosity and the flow Reynolds number.
It results that the permeability strongly affects the flow pattern. For all Reynolds numbers and thickness-to-height-ratios investigated, the baseflow is characterized by a recirculation region that gradually becomes smaller and detaches from the body as the value of $Da$ is increased, up to a critical value beyond which the recirculation region vanishes. It is also observed that the drag coefficient rapidly decreases as Da is increased further after that the recirculation bubble is disappeared. This behavior has been explained in terms of modification of the vorticity field due to the suction and blowing effect from the body walls: when the permeability increases, the wake vorticity decreases yielding a reduction of the induced counter velocity in the wake leading, in turn, to a smaller and weaker recirculation length in the streamwise direction.

The position and the elongation of the recirculation region, both of which depend on the body porosity and on the flow Reynolds number, are seen to affect the stability properties of the baseflow. In particular, for a sufficiently low value of $Da$, two complex-conjugate global modes associated with the vortex shedding become unstable when the flow Reynolds number exceeds a critical value. As Re is further increased, depending on the value of Da, the flow can become stable again. Moreover, if Re is kept constant and Da is varied, for each value of Re. A critical Darcy number $Da_{cr}$ exists beyond which the recirculation region vanishes and the flow becomes stable.
It has been observed that the marginal stability curve in the $Da-Re$ plane well correlates with one iso-level of $\Delta L$, i.e. the length of the  recirculation region. Interestingly, the flow can be unstable even without a recirculation region if sufficiently elongated region exists with a wake defect larger than $95\%$. This behaviour has been readily explained by investigating the local stability properties of the baseflow and by identifying the region of absolute instability in the wake. This analysis shows that when the velocity on the symmetry axis is less than the 5\% of the free-stream velocity the baseflow can sustain locally absolutely unstable perturbations.
The baseflow patterns and their stability properties are seen to only dependent weakly on the porosity of the body for a given permeability. On the other hand, when the thickness-to-height ratio $t/d$ is increased the corresponding stability curves are shifted towards increasing Darcy numbers meaning that higher permeability is needed to stabilize the wake past bodies which are progressively more elongated in the streamwise direction. It has also been observed that the neutral stability curves collapse on each other when they are scaled using a Darcy number based on the body's cross-section. {\lorenzo{Moreover, the existence of a general critical permeability that ensure the suppression of oscillating wakes for all the cases here presented is of fundamental importance and we expect that this feature will be identified also for wakes of other bluff bodies that show similar bifurcation scenarios than the one here investigated.}}
 \\
\indent It can be concluded that the body porosity has a significant impact not only on the baseflow configurations, as already pointed out in the literature, but also on its stability properties. Some flow patterns have been observed which are different from the usual picture we have thinking about wakes past impervious bluff bodies. These are for instance separation regions that are detached from the body and from which vortex shedding takes place, or vortex shedding originating from regions of flow which is slowed by the porous body but not recirculating and positioned downstream of the body. For all these unusual cases we have provided here a full characterisation and explanation, using numerical simulation, local and global stability analysis.

From the analysis described here it is clear that the permeability can be an effective control to stabilize the wake past porous bluff bodies. Permeability can be the result evolution in nature, as it is probably the case for particular seeds as described in the introduction. In this respect, the methods and the results provided here could help in understanding possible optimisation criteria that nature has pursued by evolution, which is a very important topic in research. Moreover, results here presented suggest how permeability can be designed on purpose in specific engineering applications for flow control, at least for what concerns wakes past plane bluff bodies. Finally the effects of permeability, that have been unraveled here in the specific case of rectangular cylinders, have potential impact in many other flow cases of interest, both from the fundamental and the practical viewpoints. 

\appendix

\section{Mathematical model for the flow inside the porous medium} 
\label{sec:app1}
The body is assumed to be a homogeneous and isotropic porous medium which characteristics are its porosity $\phi$ and permeability $k$.
Referring to a representative volume of the porous medium $V$ (Fig. \ref{fig:flowConf}a), we define the \textit{superficial average velocity} $\langle \textbf{u}_b\rangle |_{\bm{x}}=\frac{1}{V} \int_{V_f} \textbf{u}_b(\bm{x}+\bm{x_f}) \,d\Omega$
and the \textit{intrinsic average velocity}
 $\langle \textbf{u}_b\rangle_\beta |_{\bm{x}}=\frac{1}{V_f} \int_{V_f} \textbf{u}_b(\bm{x}+\bm{x_f}) \,d\Omega$, and the same definitions are valid for the pressure. Using an averaging technique, i.e. $\textbf{u}_b=\langle  \textbf{u}_b\rangle_{\beta} + \textbf{u}_b^{\prime}$ (see \cite{Ochoa-Tapia1995,Whitaker1996}), the equations for the flow through a porous medium can be written as:
\begin{equation}
\nabla \cdot \langle \textbf{u}_b\rangle =0,
\end{equation}
\begin{multline}
\frac{\rho}{\phi}\frac{\partial{\langle \textbf{u}_b\rangle }}{\partial{t}}+\frac{\rho}{\phi^2}\langle  \textbf{u}_b\rangle \cdot \nabla \langle  \textbf{u}_b\rangle  +\underbrace{\frac{\rho}{\phi} \nabla \cdot\langle \textbf{u}_b^{\prime} \cdot \textbf{u}_b^{\prime}\rangle  }_{I-subfilter \ scale \ stress} =\\
= -  \nabla\langle  p_b \rangle _{\beta} +\frac{ \mu}{\phi} \nabla ^2\langle  \textbf{u}_b\rangle  -\underbrace{\mu /k  \langle \textbf{u}_b \rangle }_{II -Darcy \ term}-\underbrace{\textbf{F}\mu /k \langle \textbf{u}_b\rangle }_{III-Forchheimer \ term}
\end{multline}

In this problem we can neglect, without effects on the flow behaviour, the terms (I) and (III). For what concerns the term (I), it is an additional contribution to diffusion, usually called \textit{mechanical dispersion}. Using a similarity with turbulent stresses (see \cite{Breugem2004}):
\begin{equation}
\frac{\rho}{\phi} \langle \textbf{u}_b^{\prime} \cdot \textbf{u}_b^{\prime}\rangle_{ij} =-\mu_{mech}( \frac{\partial\langle u_{b_{i}}\rangle }{\partial x_{j}} + \frac{\partial\langle u_{b_{j}}\rangle }{\partial x_{i}})
\end{equation}

We can estimate the mechanical viscosity using $\mu_{mech}=c_{g} l_{\beta} \sqrt{e}$, where   $e=(\langle u_{b_{i}}^{\prime}u_{b_{i}}^{\prime}\rangle )/{2} \sim \langle u_{b_{i}}\rangle^2/2$, $l_{\beta}$ is the microscopic characteristic length ($l_{\beta} \sim \sqrt{k}$), and $c_{g}$ is a coefficient that depends on the pores geometry.

Using as reference quantities the incoming velocity $U$ and the height of the rectangle $d$, we define the Reynolds number $Re=\rho Ud/\mu$ and the Darcy number $Da=k/d^2$. 

Defining the Reynolds number $Re_{\beta}$ based on the microscopic characteristic length
 $Re_{\beta}=\frac{\rho\langle {u}_b\rangle _{\beta}l_{\beta}}{\mu}  \sim \frac{ \langle {u}_b\rangle _{\beta}}{U} Re Da^{1/2}$, 
where $\frac{ \langle {u}_b\rangle _{\beta}}{U} \sim 10^{-1}$,
we can evaluate the ratio between the mechanical dispersion and the Darcy term,

\begin{equation}
R_m=\frac{ \frac{\rho}{\phi^2} \nabla \cdot\langle \textbf{u}_b^{\prime} \cdot \textbf{u}_b^{\prime}\rangle _{_i}  }{{\mu /k  \langle {u}_{b_i} \rangle} } \sim \frac{\rho \sqrt{k} |\langle \textbf{u}_b\rangle | \langle u_{b_{i}}\rangle }{d^2} \frac{k}{\mu} \sim Re_{\beta} Da  
\end{equation}

As an example, $Re=200$ and $Da=10^{-3}$; so $R_m \sim 10^{-4}$; so we can neglect the subfilter scale stress.

For what concerns the Forchheimer term, according to \cite{Whitaker1996} $F \sim c_F \cdot Re_{\beta}$, with $c_F \sim 10^{-2}$; the order of magnitude of the ratio between the Forchheimer term and the Darcy term is $10^{-3}$, and so we neglect also this term. In order to verify this assumption, some simulations have been performed (which are not reported here for the sake of brevity): an appreciable effect on the results is not observed; in particular, the variation of the pressure drop in the body is around the $1\%$, for $Re=200$ and $Da=10^{-3}$ .

The equations are made nondimensional using the incoming velocity $U$ and the height of the rectangle $d$ and they can be written as follows:

\begin{itemize} 
\item Pure fluid:
\begin{subequations}
\label{DNSfinal12}
\begin{equation}
\nabla \cdot \tilde{\textbf{u}}=0
\end{equation}
\begin{equation}
 \frac{\partial{\tilde{\textbf{u}}}}{\partial{\tilde{t}}}+\tilde{\textbf{u}}\cdot \nabla  \tilde{\textbf{u}} =
 -  \nabla \tilde{p} +\frac{1}{Re} \nabla ^2 \tilde{\textbf{u}}
\end{equation}
\end{subequations}
\item Porous medium:
\begin{subequations}
\label{DNSfinal22}
\begin{equation}
\nabla \cdot \langle\tilde{\textbf{u}}_b\rangle=0
\end{equation}
\begin{equation}
\frac{1}{\phi}\frac{\partial{\langle\tilde{\textbf{u}}_b\rangle}}{\partial{\tilde{t}}}+\frac{1}{\phi^2}\langle\tilde{\textbf{u}}_b\rangle \cdot \nabla \langle \tilde{\textbf{u}}_b\rangle =
 -  \nabla \langle\tilde{p}_b\rangle _{\beta} +\frac{1}{\phi Re} \nabla ^2 \langle\tilde{\textbf{u}}_b\rangle -\frac{1}{ReDa}  \langle\tilde{\textbf{u}}_b\rangle
\end{equation}
\end{subequations}
\end{itemize}
where the superscripts $~\tilde{\cdot}~$ represent the nondimensional quantities.

\section{Results of global stability analysis obtained using various meshes for the cylinder with t/d=0.25}
\label{sec:app2}
\begin{table}[b]
\centering
\begin{tabular}{ccccccccccccc} 
   Mesh & $x_{-\infty}$&  $x_{+\infty}$ &$y_{\infty}$&$n_L$ & $n_C$ & $n_1$ & $n_2$ & $n_3$ &  $n_s$ &$n_t$ & $10^{3}\lambda$ & $\omega$ \\ \hline
 M1&-50 &75& 40&160  & 120 & 6.3 & 4.2& 3.1& 0.4 & 86798&-1.98711& 0.70191 \\ 
 M2&-25 &50& 20&160  & 120& 6.3 & 4.2 & 3.1&0.8& 43910&-1.64652 & 0.70959\\
  M3&-25 &50& 20&160&120 &  9.6 & 7.7 & 3.8  &1.1&81370&-1.73814 & 0.70984\\
  M4 &-25 &50& 20 &160& 120&  12.5& 8.3 &6.2 &1.5&131438&-1.78355 & 0.70973\\ 
  M5 &-25 &50& 20&160 & 120 &  15 &10 & 7.4  &1.8&169862&-1.81198 &0.70984\\ \hline
\end{tabular}
\caption{Results of the mesh convergence for the configuration $t/d=0.25$, $Re=160$ and $Da=9 \times 10^{-4}$. Referring to Fig. \ref{fig:comput}, the characteristic parameters of the meshes are: $x_{-\infty}$,  $x_{+\infty}$ and $y_{\infty}$ represent the coordinates of the computational domain, respectively; $n_L$ and $n_C$ designate the vertex densities on the vertical and horizontal edge of the cylinder; $n_1$, $n_2$, $n_3$ and $n_s$ label the vertex densities on the different regions of refinement of the computational domain; $n_t$ is the total number of the elements of the grid. $\lambda$ and  $\omega$ are, respectively, the real and imaginary part of resulting global eigenvalues.
}
\label{tab:conv}
\end{table}
In this section, the effect of the spatial extent of the computational domain and vertex densities on the results of the global stability analysis is presented. The vertex densities are here controlled using different regions of refinement in the computational domain (Fig. \ref{fig:comput}). \\
The results of the mesh convergence are reported in Table \ref{tab:conv}, for the case $t/d=0.25$, $Re=160$ and $Da=9 \times 10^{-4}$ for five different meshes, denoted $M1$ to $M5$. The meshes $M1$ and $M2$ differ only for the size of the computational domain. In particular, for $M1$,  $x_{-\infty}=-50$, $x_{\infty}=75$ and $y_{\infty}=40$, whereas for $M2$,  $x_{-\infty}=-25$, $x_{\infty}=50$ and $y_{\infty}=20$ (Fig. \ref{fig:comput}). Comparing the leading global eigenvalues obtained for these two meshes, it is clear that the domain size has a negligible impact on the results, at least in the range of the parameters here considered. 
Thus, keeping constant the domain size of $M2$, the vertex density is progressively increased in the meshes $M3$, $M4$ and $M5$. The corresponding results show that also the the vertex densities have a small impact on the global stability results and, in particular, three significant digits remain constant for all the computations here performed.\\Summarizing, the results of the convergence analysis show that the spatial discretization employed in $M2$ is suitable to ensure the reliability of the results of the global stability analysis and, thus, it has been chosen to present all the results reported in the paper. 
\bibliographystyle{unsrt}
\bibliography{porous}

\begin{thebibliography}{10}

\bibitem{Brinkman1949}
H.~C. Brinkman.
\newblock A calculation of the viscous force exerted by a flowing fluid on a
  dense swarm of particles.
\newblock {\em Flow, Turbulence and Combustion}, 1(1):27, Dec 1949.

\bibitem{Sunada2002}
S.~Sunada, H.~Takashima, T.~Hattori, K.~Yasuda, and K.~Kawachi.
\newblock Fluid-dynamic characteristics of a bristled wing.
\newblock {\em Journal of experimental biology}, 205(17):2737--2744, 2002.

\bibitem{Cummins2017}
C.~Cummins, I.~M. Viola, E.~Mastropaolo, and N.~Nakayama.
\newblock The effect of permeability on the flow past permeable disks at low
  reynolds numbers.
\newblock {\em Physics of Fluids}, 29(9):097103, 2017.

\bibitem{mcginley1989fruit}
M.A. McGinley and E.J. Brigham.
\newblock Fruit morphology and terminal velocity in tragopogon dubious (l.).
\newblock {\em Functional Ecology}, pages 489--496, 1989.

\bibitem{prandtl1904verhandlungen}
L.~Prandtl.
\newblock Verhandlungen des dritten internationalen mathematiker-kongresses.
\newblock {\em Heidelberg, Leipeizig}, pages 484--491, 1904.

\bibitem{castro71}
I.~P. Castro.
\newblock Wake characteristics of two-dimensional perforated plates normal to
  an air-stream.
\newblock {\em Journal of Fluid Mechanics}, 46(3):599--609, 1971.

\bibitem{zong_nepf_2012}
L~Zong and H~Nepf.
\newblock Vortex development behind a finite porous obstruction in a channel.
\newblock {\em Journal of Fluid Mechanics}, 691:368--391, 2012.

\bibitem{jue2004numerical}
T-C. Jue.
\newblock Numerical analysis of vortex shedding behind a porous square
  cylinder.
\newblock {\em International Journal of Numerical Methods for Heat \& Fluid
  Flow}, 14(5):649--663, 2004.

\bibitem{chen2008numerical}
X.~Chen, P.~Yu, SH. Winoto, and H-T. Low.
\newblock Numerical analysis for the flow past a porous square cylinder based
  on the stress-jump interfacial-conditions.
\newblock {\em International Journal of Numerical Methods for Heat \& Fluid
  Flow}, 18(5):635--655, 2008.

\bibitem{MONKEWITZ88}
P.~A. Monkewitz.
\newblock The absolute and convective nature of instability in two dimensional
  wakes at low reynolds numbers.
\newblock {\em The Physics of Fluids}, 31(5):999--1006, 1988.

\bibitem{huerre90}
P.~Huerre and Monkewitz~P. A.
\newblock Local and global instabilities in spatially developing flows.
\newblock {\em Annual Review of Fluid Mechanics}, 22(1):473--537, 1990.

\bibitem{fabre2008bifurcations}
D.~Fabre, F.~Auguste, and J.~Magnaudet.
\newblock Bifurcations and symmetry breaking in the wake of axisymmetric
  bodies.
\newblock {\em Physics of Fluids}, 20(5):051702, 2008.

\bibitem{meliga2009unsteadiness}
P.~Meliga, J-M. Chomaz, and D~Sipp.
\newblock Unsteadiness in the wake of disks and spheres: instability,
  receptivity and control using direct and adjoint global stability analyses.
\newblock {\em Journal of Fluids and Structures}, 25(4):601--616, 2009.

\bibitem{Giannetti07}
F.~Giannetti and P.~Luchini.
\newblock {Structural sensitivity of the first instability of the cylinder
  wake}.
\newblock {\em Journal of Fluid Mechanics}, 581:167--197, 2007.

\bibitem{bj67}
Gordon~S. Beavers and Daniel~D. Joseph.
\newblock Boundary conditions at a naturally permeable wall.
\newblock {\em Journal of Fluid Mechanics}, 30(1):197?207, 1967.

\bibitem{Whitaker1986}
Stephen Whitaker.
\newblock Flow in porous media i: A theoretical derivation of darcy's law.
\newblock {\em Transport in Porous Media}, 1(1):3--25, Mar 1986.

\bibitem{Whitaker1996}
S.~Whitaker.
\newblock {The Forchheimer equation: A theoretical development}.
\newblock {\em Transport in Porous Media}, 25(1):27--61, 1996.

\bibitem{Ochoa-Tapia1995}
J.~A. Ochoa-Tapia and S.~Whitaker.
\newblock {Momentum transfer at the boundary between a porous medium and a
  homogeneous fluid-I. Theoretical development}.
\newblock {\em International Journal of Heat and Mass Transfer},
  38(14):2635--2646, 1995.

\bibitem{Gallaire2017}
F.~Gallaire and P-T. Brun.
\newblock Fluid dynamic instabilities: theory and application to pattern
  forming in complex media.
\newblock {\em Phil. Trans. R. Soc. A}, 375(2093):20160155, 2017.

\bibitem{MR3043640}
F.~Hecht.
\newblock New development in freefem++.
\newblock {\em J. Numer. Math.}, 20(3-4):251--265, 2012.

\bibitem{zampogna16}
G.~A. Zampogna and A.~Bottaro.
\newblock Fluid flow over and through a regular bundle of rigid fibres.
\newblock {\em Journal of Fluid Mechanics}, 792:5--35, 2016.

\bibitem{Biancofiore11}
L.~Biancofiore and F.~Gallaire.
\newblock The influence of shear layer thickness on the stability of confined
  two-dimensional wakes.
\newblock {\em Physics of Fluids}, 23(3):034103, 2011.

\bibitem{Meliga09}
P.~Meliga, J.-M. Chomaz, and D.~Sipp.
\newblock Unsteadiness in the wake of disks and spheres: Instability,
  receptivity and control using direct and adjoint global stability analyses.
\newblock {\em Journal of Fluids and Structures}, 25(4):601 -- 616, 2009.
\newblock Bluff Body Wakes and Vortex-Induced Vibrations (BBVIV-5).

\bibitem{Breugem2004}
W.~P. Breugem.
\newblock {The influence of wall permeability on laminar and turbulent flows,
  Theory and simulations}.
\newblock {\em Doctoral dissertation, TU Delft, Delft University of
  Technology}, 2004.

\end{thebibliography}

\end{document}